\documentclass[a4paper,12pt]{article}
\usepackage{amsfonts,amsthm,amsmath,amssymb,graphicx,color,hyperref,youngtab}
\usepackage[bulletsep]{collref}
\usepackage[lmargin=3.0cm, rmargin=1.5cm,tmargin=2.50cm,bmargin=2.50cm]{
geometry}
\usepackage{mathrsfs}
\usepackage{float}
\usepackage{ulem}
\usepackage{graphicx}

\newcommand{\be}{\begin{equation}}
\newcommand{\ee}{\end{equation}}
\newcommand{\bea}{\begin{eqnarray}}
\newcommand{\eea}{\end{eqnarray}}

\newcommand{\nn}{{\nonumber\\}}
\newcommand{\Tr}{\text{Tr}}

\newcommand\beq{\begin{equation}}
\newcommand\eeq{\end{equation}}

\begin{document}


\def\gap#1{\vspace{#1 ex}}
\def\be{\begin{equation}}
\def\ee{\end{equation}}
\def\bal{\begin{array}{l}}
\def\ba#1{\begin{array}{#1}}  
\def\ea{\end{array}}
\def\bea{\begin{eqnarray}}
\def\eea{\end{eqnarray}}
\def\beas{\begin{eqnarray*}}
\def\eeas{\end{eqnarray*}}
\def\del{\partial}
\def\eq#1{(\ref{#1})}
\def\fig#1{Fig \ref{#1}} 
\def\re#1{{\bf #1}}
\def\bull{$\bullet$}
\def\nn{\\\nonumber}
\def\ub{\underbar}
\def\nl{\hfill\break}
\def\ni{\noindent}
\def\bibi{\bibitem}
\def\ket{\rangle}
\def\bra{\langle}
\def\vev#1{\langle #1 \rangle} 
\def\lsim{\stackrel{<}{\sim}}
\def\gsim{\stackrel{>}{\sim}}
\def\mattwo#1#2#3#4{\left(
\begin{array}{cc}#1&#2\\#3&#4\end{array}\right)} 
\def\tgen#1{T^{#1}}
\def\half{\frac12}
\def\floor#1{{\lfloor #1 \rfloor}}
\def\ceil#1{{\lceil #1 \rceil}}

\def\mysec#1{\gap1\ni{\bf #1}\gap1}
\def\mycap#1{\begin{quote}{\footnotesize #1}\end{quote}}
\def\ubsec#1{\gap1\ni\underbar{#1}\gap1}

\def\Om{\Omega}
\def\a{\alpha}
\def\b{\beta}
\def\l{\lambda}
\def\m{\mu}
\def\n{\nu}
\def\om{\omega}
\def\s{\sigma}
\def\g{\gamma}
\def\t{\tau}

\def\lan{\langle}
\def\ran{\rangle}

\def\bit{\begin{item}}
\def\eit{\end{item}}
\def\benu{\begin{enumerate}}
\def\eenu{\end{enumerate}}
\def\eps{\epsilon}


\def\bT{{\bar T}}
\def\bL{{\bar L}}
\def\lt{{\tilde \lambda}}
\def\vt{{\tilde v}}
\def\wt{{\tilde w}}
\def\omt{{\tilde\omega}}
\def\ut{{\tilde u}}
\def\bP{{\bf P}}
\def\bQ{{\bf Q}}
\def\Jt{{\tilde J}}
\def\zb{{\bar z}}
\def\wb{{\bar w}}
\def\hb{{\bar h}}
\def\nor#1{{:\kern-1pt #1 \kern-1pt:}}
\def\wc{{\mathcal W}}



\rightline{TIFR/TH/15-02}
\vspace{1.2truecm}

\vspace{1pt}


{\LARGE{
\begin{center}{\bf Thermalization with chemical potentials, and
higher spin black holes}
\end{center}
}}

\vskip.9cm

\thispagestyle{empty} \centerline{\large \bf  
Gautam Mandal,\footnote{mandal@theory.tifr.res.in} Ritam Sinha,\footnote{ritam@theory.tifr.res.in} and Nilakash Sorokhaibam\footnote{nilakashs@theory.tifr.res.in}}
    
\vspace{.8cm} 
\centerline{{\it Department of Theoretical Physics}}
\centerline{{\it Tata Institute of Fundamental Research, Mumbai
    400005, India.} }


\gap7

\centerline{\today}

\gap3

\thispagestyle{empty}

\gap6

\centerline{\bf Abstract}
\vskip.5cm 

We study the long time behaviour of local observables following a
quantum quench in 1+1 dimensional conformal field theories possessing
additional conserved charges besides the energy. We show that the
expectation value of an arbitrary string of {\it local} observables
supported on a finite interval exponentially approaches an equilibrium
value. The equilibrium is characterized by a temperature and chemical
potentials defined in terms of the quenched state. For an infinite
number of commuting conserved charges, the equilibrium ensemble is a
generalized Gibbs ensemble (GGE). We compute the thermalization rate
in a systematic perturbation in the chemical potentials, using a new
technique to sum over an infinite number of Feynman diagrams.  The
above technique also allows us to compute relaxation times for thermal
Green's functions in the presence of an 
arbitrary number of chemical potentials.  In the
context of a higher spin (hs[$\l$]) holography, the partition function
of the final equilibrium GGE is known to agree with that of a higher
spin black hole. The thermalization rate from the CFT computed in our
paper agrees with the quasinormal frequency of a scalar field in this
black hole.

\setcounter{page}{0} \setcounter{tocdepth}{2}

\newpage

\tableofcontents

\section{\label{sec:intro}Introduction and Summary}

The study of thermalization in closed interacting quantum systems has
a long history (see, e.g. \cite{Polkovnikov:2010yn} for a review).  It
has been known ever since the celebrated work of Fermi, Pasta and Ulam
(FPU) that interacting classical systems need not necessarily
equilibrate. The question of finding sufficient conditions for
thermalization in quantum systems is also an open one.  Recently, the
advent of holography has linked the issue of thermalization in
strongly coupled quantum field theories to another important,
classical, problem of black hole formation (see, e.g.
\cite{Balasubramanian:2010ce, Liu:2013qca, Bhattacharyya:2009uu,
  Caceres:2014pda} and references therein). In the latter setting too,
the issue of gravitational collapse of a given matter distribution is
rather nontrivial; indeed there is an interesting debate in the
current literature (see, e.g., \cite{Bizon:2011gg, Dias:2011ss,
  Balasubramanian:2014cja, Basu:2014sia,Craps:2014vaa, Craps:2014jwa})
regarding the fate of perturbations in anti-de-Sitter spacetimes.

In this paper, we will focus on two-dimensional conformal field
theories (CFTs) on an infinite line $\s \in {\mathbb R}$. We will
consider the system at $t=0$ to be in a ``quenched 
state''\footnote{In  the original sense of the term, a quantum quench 
is defined as a sudden change from a hamiltonian $H_0$  to a
hamiltonian $H$ which governs further evolution for $t\ge 0$. The system is
assumed to be in the ground state of $H_0$ at $t=0$, which 
serves as an initial state for subsequent dynamics;
the dynamics is nontrivial since the initial state prepared 
this way is not an eigenstate of $H$. 
In this paper, as in
\cite{Calabrese:2005in}, we will mean by a ``quenched state'' simply a
pure state which is not an eigenstate of the Hamiltonian $H$. The
kind of quenched state defined in \eq{psi-ep-0} is  sometimes said to
describe 
a global quench or a homogeneous quench, as the state is translationally
invariant. We will briefly mention inhomogeneous and local quenches
in Section \ref{sec:discuss}.}
\begin{align}
| \psi_0 \ran=  \exp[-\eps_2 H -\sum_{n=3}^\infty \eps_n W_n]| Bd \ran
\label{psi-ep-0}
\end{align}
Here $| Bd \rangle$ is a conformal boundary state; the exponential
factors cut off the UV modes to make the state normalizable. $W_n$
denote the additional conserved charges in the theory.\footnote{For
  the purposes of this paper, we will identify them with $W_n$-charges
  of 2D CFT, $n=3,4,...$ (with $W_2 \equiv H$), although much of what
  we say will go through independent of this specific choice as long
  as these charges mutually commute and are defined from currents which are quasiprimary fields of the conformal algebra.}
This choice of the quenched state is a generalization of 
that in \cite{Calabrese:2005in} for which $\eps_n=0$, for $n>3$. 
 
The wavefunction for $t>0$ is given by  
\begin{align}
| \psi(t)\ran =  \exp[- i H t]| \psi_0 \ran
\label{psi-t}
\end{align}

The questions we will explore, and answer, are: what is the long time
behaviour of various observables in $|\psi(t)\ran$?  In particular,
does the expectation value of an operator (or a string of operators)
approach a constant? If so, (i) is the constant value characterized by
a thermodynamic equilibrium, and (ii) what is the rate of approach to
the constant value?  More generally, we would also address, and
partially answer, the questions: how does the existence and rate of
thermalization depend on the initial state and the choice of
observables?




\paragraph{Thermalization}
We find in this paper that the expectation values of local observables
(supported on a finite interval $A:\s \in [-l/2, l/2]$)
asymptotically approach (see \eq{string-infinity} for the precise
statement) their values in an equilibrium ensemble,
\begin{align}
&\rho_{eqm} =  \frac1Z  \exp[-\b H -\sum_n \mu_n W_n],\kern10pt
Z= \Tr  \exp[-\b H -\sum_n \mu_n W_n]
\label{ensemble}
\end{align}
whose temperature and chemical potentials are related to the cutoff
scales in \eq{psi-0} as follows
\begin{align}
\b = 4 \eps_2, \;\;  \mu_n = 4 \eps_n, \;\; n=3,4,... 
\label{eq-ensemble}
\end{align}
The relations \eq{eq-ensemble} are uniquely dictated by the
requirement that the expectation values of the conserved charges $H,
W_3, W_4,...$ in the initial state match those in the mixed state \eq{ensemble}
(see \eq{charge-match}). In the absence of the extra parameters
$\eps_n, n=3,4,...$ this result is derived by the elegant method of
conformal transformations \cite{Calabrese:2005in}. In the presence of
these parameters, this method is not available; in this paper, we deal
with the extra exponential factors in terms of an infinite series and
do a resummation.

We emphasize that the thermalization we found above persists even when
we have an integrable model with an infinite number of conserved
charges. Relaxation in integrable systems has been found in recent
years in the context of, e.g., (a) one-dimensional hardcore bosons
\cite{Rigol:2007}, (b) transverse field Ising model
\cite{Calabrese:2011GGE}, and (c) matrix quantum mechanics models
\cite{Mandal:2013id}. The equilibrium ensembles in this context have
been called a generalized Gibbs ensemble (GGE). Our present result on
integrable conformal field theories adds to the list of these
examples. Interestingly, the thermalization we find works even for
free conformal field theories, e.g. a free scalar field
theory.\footnote{\label{ftnt:caldeira}This happens essentially due to
  the fact that we consider here thermalization of local observables
  and that local field modes are mutually coupled even in a free field
  theory. Thermalization happens at times greater than the scale of
  localization, as we will see below.}

With the above identification of parameters, we will rewrite the
initial quenched state \eq{psi-ep-0} henceforth as
\begin{align}
| \psi_0 \ran=  \exp[-(
\b H -\sum_{n=3}^\infty \mu_n W_n)/4]| Bd \ran
\label{psi-0}
\end{align}

We find the following specific results:


\paragraph{1. Thermalization time scale for single local observables:}
 
We find that at large times
\begin{align}
\lan \psi(t)| \phi_k(\s)| \psi(t) \ran = 
\Tr\left( \phi_k(0) \rho_{eqm}(\b, \mu_i)\right) 
+  a_k\, e^{-\g_k t}  + ...
\label{one-pt}
\end{align}
where $\phi_k(\s)$ is an arbitrary quasiprimary field (labelled by an
index $k$).  Below we compute the thermalization exponent $\g_k$ in a
perturbation in the chemical potentials and to linear order it is
given by
\begin{align}
&\g_k = \frac{2\pi}{\b} \left[\Delta_k + \sum_n 
\tilde\mu_n  Q_{n,k} + O({\tilde\mu}^2)\right],
\kern3pt
\tilde \mu_n \equiv \frac{\mu_n}{\b^{n-1}},
\label{gamma-k}
\end{align}
Here $\Delta_k =h_k + \bar h_k$ is the scaling dimension and ${
  Q}_{n,k}$ are the (shifted) $W_n$-charges (see \eq{o-mu-n} for the
full definition) of the field $\phi_k$ (in case of primary fields) or
of the minimum-dimension field which appears in the conformal
transformation of $\phi_k$. To obtain this result, we perform the
infinite resummation mentioned below \eq{eq-ensemble}. At large times,
the perturbation series for the one-point function in the chemical
potentials exponentiates (see \eq{o-mu-n}), to give the corrected
exponent in the above equation. {\it In various related contexts,
  finite orders of perturbation terms in chemical potentials have been
  computed before
  \cite{Gaberdiel:2013jca,Beccaria:2013yca,Datta:2014uxa}.  Our
  finding in this paper is that at large times, there is a regularity
  among the various orders leading to an exponential function as in
  \eq{one-pt} (see Section \ref{sec:o-mu-all} for details).}

\noindent {\bf Universality}: In the case of zero chemical potentials,
it has been noted in \cite{Calabrese:2007-global}, that although the
relaxation time $\tau_k= \pi\eps_2/(2\Delta_k)$ $=2\pi \b/(\Delta_k)$
is non-universal (in the sense that it depends on the specific initial
state \eq{psi-ep-0}), the ratio of relaxation times for two different
fields, namely, $\tau_{k_1}/\tau_{k_2}= \Delta_{k_2}/\Delta_{k_1}$ is
universal (it depends only on the CFT data and not on the initial
state and is hence expected to be valid for a general class of initial
states). In the presence of  the
additional cut-off parameters $\eps_i, i=3,...$ in the initial state
\eq{psi-ep-0}, the ratio $\tau_{k_1}/\tau_{k_2}=$ $\g_{k_2}/\g_{k_1}$=
$ \left(\Delta_{k_2} + \sum_n \tilde\mu_n Q_{n,k_2}\right)$/$
\left(\Delta_{k_1} + \sum_n \tilde\mu_n Q_{n,k_1}\right)$ is, however,
not independent of the initial state.

However, as we will briefly discuss in Section \ref{sec:discuss}, for a large class of quench states
(e.g. where the energy density is uniform outside of a domain of
compact support) the $\b$-dependence of $\tau_k$, in
the absence of chemical potentials, can be understood as
the dependence on the uniform energy density (see a related discussion
in \cite{Gupta:2008th}).  The time scales $\tau_k$, therefore, do have a
limited form of universality in the sense that it depends on a rather
robust feature of the initial state.  Our calculations in this paper
leads us to believe that this feature will continue in the presence of
chemical potentials, in the sense that the additional dependence
of the time scales $1/\g_k$ on the $\mu_n$ is fixed by the charge
densities corresponding to the additional conserved charges. We hope to
address this in \cite{GM:2015Progress2}.

\paragraph{2. Multiple local observables, reduced density matrix:}

Besides the one-point functions discussed above, it turns our that
we can  demonstrate thermalization of {\it all
  operators in an interval $A$ of length $l$}. It is convenient to
define a `thermalization function' $I_A(t)$ \cite{Cardy:2014rqa} as
\begin{align}
&
I_A(t)= \Tr(\hat \rho_{dyn,A}(t) \hat \rho_{eqm,A}(\b, \mu_n))=
  \frac{\Tr(\rho_{dyn,A}(t) \rho_{eqm,A}(\b, \mu_n))}
       {\left[\Tr(\rho_{dyn,A}(t)^2)\Tr (\rho_{eqm,A}(\b,
           \mu_i)^2)\right]^{1/2}} 
\nonumber\\ &\rho_{dyn, A}(t)=
       \Tr_{\bar A}\ | \psi(t) \ran \lan \psi(t) |, \kern3pt
       \rho_{eqm, A}(\b, \mu_n)= \Tr_{\bar A}\ \rho_{eqm}(\b, \mu_i)
\label{I-t}
\end{align}
Here $\hat \rho = \rho/\sqrt{\Tr \rho^2}$ denotes a
`square-normalized' density matrix.\footnote{\label{ftnt:vector}Note
  that operators in a Hilbert space ${\mathsf H}$ can themselves be
  regarded as vectors in ${\mathsf H} \times {\mathsf H}^*$; under
  this interpretation $\Tr(A\ B)$ defines a positive definite scalar
  product. With this understanding, we will regard the hatted density
  matrices as unit vectors.}\footnote{Throughout this paper, we will
  consider field theories with an infinite spatial extent. The entire
  Hilbert space is assumed to be of the form ${\mathsf H}_A \otimes
  {\mathsf H}_{\bar A}$. $\Tr_{\bar A}$ implies tracing over ${\mathsf
    H}_{\bar A}$.}  We show below that at large times the
thermalization function has the form
\begin{align}
&I_A(t)= 1- \a(\tilde l)\  e^{-2 \g_m t} + ..., \kern3pt  
\tilde l \equiv l/\b
\label{I-asymp}
\end{align}
where $\g_m$ refers to the exponent \eq{gamma-k} for the operator
$\phi_m$ with minimum scaling dimension.\footnote{We will assume here
  that the spectrum of such $\Delta$'s is bounded below by a finite
  positive number. In case of a free scalar field theory, we can
  achieve this by considering a compactified target space.} $\a(\tilde
l)$ is computed as a power series in $\tilde l$ which we find using
the short interval expansion, valid for $\tilde l \ll 1$, i.e. $l\ll
\b$.


Two immediate consequences of \eq{I-asymp} are
\begin{enumerate}
\item[(i)] \underbar{Thermalization of an arbitrary string
of operators:} 
Note, from \eq{I-asymp}, that
\begin{align}
I_A(t) \xrightarrow{t\to \infty} 1, 
\label{I-infinity}
\end{align}
Since the square-normalized density matrices can be regarded
as unit vectors (in the sense of footnote \ref{ftnt:vector}), 
and $I_A(t)$ can be regarded as the scalar product 
$\hat \rho_{dyn,A}(t){\bf \cdot} \hat \rho_{eqm,A} $, 
\eq{I-infinity} clearly implies 
\begin{align}
\hat \rho_{dyn,A}(t) \xrightarrow{t\to \infty} \hat
  \rho_{eqm,A} 
\label{rhoA-infinity}
\end{align}
This implies the following statement of thermalization for an
arbitrary string of local operators (with $ \s_1, \s_2,...  \in A$)
\begin{align}
\lan \psi(t) | O(\s_1,t_1) O(\s_2, t_2) ...| \psi(t) \ran 
&=  \Tr( \hat \rho_{dyn,A}(t) O(\s_1,t_1) O(\s_2,t_2) ...  )
\nonumber\\
& \kern60pt \xrightarrow{t\to \infty}
\Tr( \hat \rho_{eqm,A} O(\s_1,t_1) O(\s_2,t_2) ...  ).
\label{string-infinity}
\end{align}

\item[(ii)] \underbar{Long time behaviour of reduced density matrix:}

Carrying on with the interpretation of $I_A(t)$ as a scalar product,
we can infer following asymptotic behaviour of $\hat \rho_{dyn}(t)$
from \eq{I-asymp}:
\begin{align}
\hat \rho_{dyn,A}(t) = \hat \rho_{eqm,A}(\b, \mu_i) & \left( 1-
  \a\, e^{-2 \g_m t } +  ...\right) 
+ \hat Q \left(\sqrt{2 \a}\, e^{-  \g_m t } + ... \right)
\label{q-hat}
\end{align} 
where $\Tr({\hat Q}^2)=1,\;\; \Tr( \hat \rho_{eqm,A}(\b, \mu_i) \hat Q
)=0. $ We will specify further properties of $\hat Q$ later on.

{\bf Importance of local observables:} In case of a free massless
scalar field, it is easy to show that quantities like $\lan \psi(t)|
\a_1^2 \a_1^\dagger |\psi(t) \ran$ perpetually oscillate and never
reach a constant (see a related calculation in \cite{Mandal:2013id}).
The modes $\a_n$ represent Fourier modes and are non-local. Indeed, as
\cite{Cardy:2014rqa,Calabrese:2006quench,Calabrese:2007-global}
showed, in the absence of chemical potentials, the exponential term in
\eq{I-asymp} is $e^{-2\g_m(t- l/2)}$ and the thermalization sets in
only after $t$ exceeds $l/2$. Thus, for $l=\infty$, there is no
thermalization, which is consistent with the above observation about
perpetual oscillations. We expect the form $e^{-2\g_m(t- l/2)}$ to
continue to hold in the presence of chemical
potentials\footnote{Although, in the short-interval expansion employed
  in this paper to derive \eq{I-asymp}, which uses $t\ gg \b \gg l$,
  such an $l$-dependence in the exponent cannot be easily seen from
  the pre-factor $\a(\tilde l)$ unless one sums over an infinite
  orders in $\tilde l$.}, since the effect of the chemical potentials
on the exponent $\g_k$ can be viewed as a shift of the anomalous
dimension $\delta \Delta_k = \sum_n \tilde\mu_n Q_{n,k} +
O({\tilde\mu}^2)$ (see, e.g. \eq{cft-qnm-beta-mu}). This shows that,
as in the case of zero chemical potentials, equilibration sets in
only after $t$ exceeds $l/2$. We will see a similar phenomena next in
the context of a decay of perturbations to a thermal state.

\end{enumerate}

\paragraph{3. Decay of perturbations to a thermal state:} 
We compute (see Section \ref{sec:qnm-cft} for details) the
time-dependent two-point Green's function $G_+(t,l;\b,\mu)$ for two
points spatially separated by a distance $l$. We find that for $t, l,
t-l \gg \b$, the time-dependence is exponential, with the same
exponent as in \eq{one-pt}:
\begin{align}
G_+(t,l;\b,\mu)  \equiv \frac1{Z} 
\Tr\left(\phi_k(l,t) \phi_k(0,0) e^{-\b H -  \sum_n \mu_n W_n}\right)
=\hbox{const}~ e^{- \g_k t}
\label{thermal-decay-mu}
\end{align}  
Note that the above thermalization sets in for $t>l$. For $t< l$, the
two-point function has an exponential decay in the spatial separation
(see Section \ref{sec:qnm-cft} and Figure \ref{thermal-fig}).

The computation of the above relaxation times in the presence of an
arbitrary number of chemical potentials uses the technique, described
above, of summing over an infinite number of Feynman diagrams, and is
one of the main results of our paper.

\paragraph{4. Collapse to higher spin black holes:} 

In \cite{Maldacena:2001kr, Hartman:2013qma} the bulk dual to the
time-dependent state \eq{psi-t} corresponding to initial condition
\eq{psi-0}, for large central charges, has been constructed in the
case of zero chemical potentials. The dual geometry corresponds to one
half of the eternal BTZ (black string) geometry, whose boundary
represents an end-of-the-world brane. In \cite{Caputa:2013eka} the
result has been extended to the case of non-zero angular momentum and
a Chern-Simons charge.  In case of an infinite number of chemical
potentials, a bulk dual to the equilibrium ensemble \eq{ensemble} has
been identified, in the context of the Gaberdiel-Gopakumar hs$(\l)$
theory \cite{Gaberdiel:2010pz}, as a higher spin black hole with those
chemical potentials \cite{Gutperle:2011kf,Kraus:2011ds}. It is natural
to conjecture \cite{Caputa:2013eka,Mandal:2013id} that the
time-development \eq{psi-t} should be dual to a collapse to this
higher spin black hole.  At late times, therefore, the thermalization
exponent found above should correspond to the quasinormal frequency of
the higher spin black hole. We find that (see Section \ref{sec:qnm}
and \cite{GM:2015Progress1}) this is indeed borne out in a specific
example.

\gap3

The plan of the paper is as follows. The results 1, 2, 3 and 4 above are
described in Sections \ref{sec:one-pt}, \ref{sec:i-t},
\ref{sec:qnm-cft} and \ref{sec:qnm}, respectively. The resummation of
an infinite number of Feynman diagrams (corresponding to insertions of
arbitrary number of chemical potential terms) is discussed in Section
\ref{sec:o-mu-all}, which uses results in Appendix
\ref{app:one-pt}. The calculation of the overlap of reduced density
matrices in Section \ref{sec:i-t} needs the use of the short-interval
expansion, which is described in Appendix \ref{sec:short}. In Section
\ref{sec:discuss} we present our conclusions and make some remarks on
inhomogeneous quench \cite{GM:2015Progress2}.

\section{\label{sec:one-pt}One-point functions}

In this section we will consider the behaviour of the following
one-point functions of a quasiprimary field $\phi_k(\s)$
\begin{align}
& \lan\phi_k(\s,t) \ran_{dyn}
 \equiv \lan \psi(t)| \phi_k(\s)| \psi(t) \ran,
\nonumber\\
&  \lan\phi_k(\s) \ran_{eqm} \equiv  \Tr\left(
\phi_k(\s) \rho_{eqm}(\b, \mu_n)\right)
\label{one-pts}
\end{align} 
We will briefly recall how these are computed in the absence of the
chemical potentials \cite{Calabrese:2005in,DiFrancesco:1997nk}. The
first expectation value corresponds to the one-point function on a
strip geometry, with complex coordinate $w= \s + i \t$, $\s\in
(-\infty, \infty)$, $\t \in (-\b/4, \b/4)$ where $\t$ is eventually to
be analytically continued to $\t= it$. This can be conformally
transformed to an upper half plane by using the map
\begin{align}
z= i e^{(2\pi/\b) w}
\label{map}
\end{align}
For a primary field  with $h_k= \hb_k$ (of the form 
$\phi_k(w,\bar w)= \varphi_k(w)\varphi_k(\bar w)$), this procedure gives
\footnote{The subscripts {\it str, cyl} will denote a `strip' and a
  `cylinder', respectively.}(for other primary fields, the one-point
function vanishes)
\begin{align}
\lan\phi_k(\s,t) \ran_{dyn} &=
\lan \phi_k(w,\wb)\ran_{str} = \left(\frac{\del z}{\del
    w}\right)^{h_k} \left(\frac{\del \zb}{\del \wb}\right)^{\hb_k}
  \lan \phi_k(z,\zb) \ran_{UHP} 
\nonumber\\
& = a_k \left(e^{2\pi t/\b} + e^{-2\pi t/\b}\right)^{-2h_k}
\sim a_k e^{-\g^{(0)}_k t} + ..., \kern5pt \g^{(0)}_k= 2\pi \Delta_k/\b= 
4\pi h_k/\b 
\label{one-pt-prim-0}
\end{align}
We have used the following result for the one-point
function on the UHP:
\begin{align}
& \lan \phi_k(z,\zb) \ran_{UHP}
= A_k \lan \varphi_k(z) \varphi_k^*(z') \ran_{UHP}
=  A_k (z- z')^{-2 h_k},\;\;  h_k= \hb_k, \; z'=\zb
\label{images}
\end{align}
which follows by using the method of images where the antiholomorphic
factor of $\phi_k(z,\zb)$ on the upper half plane at the point
$(z,\zb)$ is mapped (up to a constant) to the holomorphic
$\varphi^*_k$ \footnote{We distinguish $\varphi^*_k$ from $\varphi_k$
  to allow for charge conjugation.} on the lower half plane at the
image point ($z', \zb'$) with $z'=\zb, \zb'= z$
\cite{Cardy:1984bb,DiFrancesco:1997nk}. In the above $a_k, A_k$ are
known numerical constants. Note that
\begin{align}
z=  i e^{2\pi(\s + i \t)/\b}=  i e^{2\pi(\s-t)/\b},\;\;
z'=\zb=  -i e^{2\pi(\s - i \t)/\b}=  -i e^{2\pi(\s+t)/\b}  
\label{image-pt}
\end{align}
so that in the large time limit we have 
\begin{align}
t\to \infty \Rightarrow z \to 0, \zb \to -i \infty.
\label{long-time}
\end{align}

The second, thermal, expectation value in \eq{one-pts}, for $\mu_n=0$,
corresponds to a cylindrical geometry in the $w$-plane, with $\t=0$
identified with $\t=\b$. By using the same conformal map \eq{map} this
can be transformed to a one-point function on the plane. For a primary
field the latter vanishes. Hence \eq{one-pt} is trivially satisfied.

For a quasiprimary field $\phi_k$, its conformal transformation
generates additional terms, including possibly a c-number term $c_k$
(e.g.  the Schwarzian derivative term for $\phi_k=T_{ww}$) and
generically lower order operators. The c-number term does not
distinguish between a plane and an UHP. This leads to the following
overall result (for $\mu_n=0$):
\begin{align}
&\lan\phi_k(\s) \ran_{eqm}=\lan \phi_k(w,\wb)\ran_{cyl} = c_k, 
\nonumber\\ 
&\lan\phi_k(\s,t) \ran_{dyn}=
\lan \phi_k(w,\wb)\ran_{str} = c_k + a_k e^{-\g_k^{(0)} t} + ..., \kern5pt
  \g_k^{(0)}= 2\pi \Delta_k/\b,
\label{one-pt-0}
\end{align}
where $\Delta_k$ now is the scaling dimension of the minimum-dimension
operator in a $T(z_1)\phi_k(z)$ OPE. This is clearly of the general
form \eq{one-pt} for $\mu_n=0$.

\gap3

We now turn to a discussion of these expectation values \eq{one-pts}
in the presence of chemical potentials $\mu_n, n=3,4,...$, as in
\eq{psi-0} and \eq{ensemble}. We will denote the new conserved
currents as $\wc_n(w)$ and $\bar\wc_n(\wb)$, $n=3,4,...$.  The
conserved charge, $W_n$,  is defined as
\begin{align}
& W_n= \frac1{2\pi} \int_\Gamma W_{\tau\tau...\tau} d\sigma 
=\frac1{2\pi}  
\int_\Gamma\left( i^n  
dw_1\, \wc_n(w_1)  + (-i)^n  d\wb_1\, \bar\wc_n(\wb_1)  
\right)
\label{contour}
\end{align}   
Here the contour $\Gamma$ is taken to be a $\t=$ constant line along
which $dw_1= d\wb_1= d\s$. Under the conformal transformation \eq{map}
to the plane/UHP, the holomorphic part of the contour integral becomes
\begin{align}
W_n|_{hol}= 
\frac{i^n}{2\pi} \left(\frac{2\pi}{\b}\right)^{n-1}
\int_{\Gamma_1}\, dz_1\,\left[  z_1^{n-1}  \wc_n(z_1) + 
\sum_{m=1}^{\floor{n/2}} a_{n,n-2m}z_1^{n-2m-1}  \wc_{n-2m}(z_1) \right]
\label{contour-z}
\end{align}   
where the $a_{n,n-2m}$ denote the mixing of $\wc_n(z_1)$ with lower
order $W$-currents under conformal transformations \cite{Pope:1991ig,
  Bouwknegt:1992wg}. The contour $\Gamma_1$ is an image of the contour
$\Gamma$ onto the plane. The expression for the antiholomorphic part
$W_n|_{antihol}$ is similar.

As mentioned before, in this paper we will regard the $W_n$ as
conserved charges of a W-algebra, although the results we derive will
be equally valid as long as these charges, together with $H$, form a
mutually commuting set, and the currents ($\wc_n(w),\bar\wc_n(\wb)$)
are quasiprimary fields.

\subsection{\label{sec:one-pt-cyl}One-point function on the cylinder 
with chemical potentials}

For simplicity we first consider the equilibrium expectation value in
\eq{one-pts}. Unfortunately, unlike the thermal factor above, the
factor $e^{-\sum_n \mu_n W_n}$ in \eq{ensemble} cannot be dealt with
in terms of a conformal map. We will, therefore, treat this factor as
an operator insertion, and write
\begin{align}
\lan\phi_k(\s) \ran_{eqm} \equiv \Tr\left(\phi_k(w,\bar w) \rho_{eqm}(\b,
\mu_n)\right)
=\frac{\lan e^{-\sum_n\mu_n W_n}\phi_k(w,\bar w)\ran_{cyl}} {\lan
  e^{-\sum_n \mu_nW_n} \ran_{cyl}}
\equiv \lan \phi_k(w,\wb)\ran_{cyl}^{\mu}
\label{cyl-mu}
\end{align} 
We will now illustrate how to compute this for a single chemical
potential, say $\mu_3$, using perturbation theory Feynman
diagrams:\footnote{The superscript {\it conn} denotes `connected'.}
\begin{align}
 \lan
 \phi_k(w,\wb)\ran_{cyl}^{\mu}=\lan\phi_k(w,\wb)\ran_{cyl}-\mu_3
 \lan W_3\phi_k(w,\wb)\ran_{cyl}^{conn} + \frac{\mu_3^2
 }{2!}\lan\ W_3 W_3 \phi_k(w,\wb)\ran_{cyl}^{conn}
 +\mathcal{O}(\mu_3^3)
\label{one-pt-fn-cyl}
\end{align}
The first term in the above expression is the constant $c_k$ that we
already encountered in \eq{one-pt-0}. For a holomorphic primary field
$\phi_k$, the second, $O(\mu_3)$, term, transformed on to the plane, gives
\begin{align}
 \lan W_3\phi_k(w)\ran_{cyl}^{conn}=\frac{2\pi}{\b^2}
 z^{h_k}\left[i^3\int_{\Gamma_1}\!\!dz_1\ z_1^2
   \lan\wc_3(z_1)\phi_k(z)\ran_{\mathbb{C}}^{conn} 
   +(-i)^3\int_{\Gamma_1}\!\! d\bar{z}_1\ \bar{z}_1^2\lan\bar{\wc}_3(\bar
   z_1)\phi_k(z)\ran_{\mathbb{C}}^{conn}\right]
\label{cyl-o-mu}
\end{align}
Here we have used the contour representations \eq{contour} and
\eq{contour-z}.  The correlator inside the second integral obviously
vanishes (it factorizes into a holomorphic and an antiholomorphic
one-point functions, leading to a vanishing connected part). The first
integral vanishes unless $\phi_k = \wc_3$ (this uses the orthogonality
of the basis of quasiprimary fields). In the latter case, using
\[
\lan\wc_3(z_1)\wc_3(z)\ran_{\mathbb{C}}=\frac{c/3}{(z_1-z)^6} 
\]
the integral evaluates to $c/(90 z^3)$; combining with the factor of
$z^3$ outside ($h_k=3$ in this case) we get a $z$-independent
constant, as we must, because of translational invariance on the
plane.  With an antiholomorphic primary field $\phi_k$, the
calculation is isomorphic. For a primary field with nonvanishing $h_k,
\bar h_k$ the result vanishes. For quasiprimary $\phi_k$, as well as
for other $W_n$ charges, the conformal transformation to the plane
additionally generates lower order operators (see, e.g.
\eq{contour-z})), each of which can be dealt with as in \eq{cyl-o-mu}.
The result is a finite constant which we will denote as
\[
\lan W_n \phi_k(w,\bar w)\ran=  c_{n,k}
\] 
(this will be non-vanishing only for special choices of $\phi_k$,
e.g. $\phi_k= \wc_n$).  As explained above, for $n=3$ and
$\phi_k(w,\bar w) = \wc_3(w)$, $ c_{n,k}= -2\pi c/(90\b^2)$.

In a similar fashion, the $O(\mu_3^2)$ term in \eq{one-pt-fn-cyl} can
be transformed to the plane. Again, we present the explicit expression
for the simple case of a holomorphic primary field $\phi_k$.
\begin{align}
&\lan W_3W_3\phi_k(w)\ran_{cyl}^{conn}=\big(\frac{2\pi}{\b^2}\big)^2 
z^{h_k}
\bigg[i^6\int_{\Gamma_1}\!\! dz_1\int_{\Gamma_2}\!\!dz_2 
\lan\wc_3(z_1)\wc_3(z_2)\phi_k(z)\ran_{\mathbb{C}}^{conn}\ z_1^2 z_2^2\nonumber\\
&+(-i)^6\int_{\Gamma_1}\!\!\!
d\bar{z}_1\int_{\Gamma_2}\!\!\!d\bar{z}_2 \lan\bar\wc_3(\bar{z}_1)
\bar\wc_3(\bar{z}_2)\phi_k(z)\ran_{\mathbb{C}}^{conn}\  \bar{z}_1^2 \bar{z}_2^2
+\int_{\Gamma_1}\!\!\!dz_1\int_{\Gamma_2}\!\!\!d\bar{z}_2 \lan\wc_3(z_1)
\bar\wc_3(\bar{z}_2)\phi_k(z)\ran_{\mathbb{C}}^{conn}\ z_1^2 \bar{z}_2^2
\nonumber\\
&+\int_{\Gamma_1}\!\!d\bar{z}_1\int_{\Gamma_2}\!\!dz_2
\lan\bar\wc_3(\bar{z}_1)\wc_3(z_2)\phi_k(z)
\ran_{\mathbb{C}}^{conn}\ \bar{z}_1^2 z_2^2\bigg]
\label{cyl-o-mu-2}
\end{align}
For holomorphic quasiprimary $\phi_k$, additional, similar, terms
appear due to the generation of lower order operators under conformal
transformation to the plane. Only the holomorphic correlator survives
(as in the O($\mu_3$) calculation). Thus, e.g. if $\phi_k=
T(z)$, the stress tensor, we have 
\[
\lan\wc_3(z_1)\wc_3(z_2){T}(z)\ran_{\mathbb{C}}=
\frac{c}{(z_1-z_2)^4(z_1-z)^2(z_2-z)^2}
\]
Again, after performing the integration over $z_1$ and $z_2$, we
obtain a $z$-independent constant, as we must. The analysis of more
general fields $\phi_k$ and two arbitrary $W$-charges is
straightforwardly generalizable. The result is a finite constant 
(can be zero for a particular $\phi_k$) which we denote as
\[
\lan W_m W_n \phi_k(w,\bar w) \ran=  c_{mn,k}
\]
Note that in \eq{cyl-o-mu-2} the result does not depend on the
location of the contours $\Gamma_1, \Gamma_2$ on the plane, since the
$W$-currents are conserved.

Summarizing, we get 
\begin{align}
\lan \phi_k(w,\wb)\ran_{cyl}^{\mu} = c_k - \sum_n \mu_n\ c_{n,k} 
+ \frac1{2!} \sum_{m,n} \mu_m \mu_n\  c_{mn,k} + O(\mu^3)
\label{one-pt-cyl-all}
\end{align} 

\subsection{\label{sec:mu-n-1pt}One-point function on the strip with 
chemical potentials}

Similarly to the previous subsection, we will treat the
$\mu$-deformations in \eq{psi-0} as operator insertions:
\begin{align}
\lan\phi_k(\s,t) \ran_{dyn} \equiv 
\lan \psi(t)| \phi_k(\s)| \psi(t) \ran
=\frac{\lan e^{-\sum_n\mu_n W_n/4}\,\phi_k(w,\bar w)\,e^{-\sum_n\mu_n W_n/4}
\ran_{str}} {\lan
  e^{-\sum_n \mu_nW_n/2} \ran_{str}}
\equiv \lan \phi_k(w,\wb)\ran_{str}^{\mu}
\label{str-mu}
\end{align} 
As before, we begin by illustrating the
calculation of this quantity with the simplest case of a single
chemical potential $\mu_3$, using perturbation theory Feynman
diagrams:
\begin{align}
&\lan \phi_{k}(w, \wb) \ran_{str}^{\mu} = \lan \phi_{k}(w, \wb) \ran_{str}
-\frac{\mu_3}{4}\lan\{W_3,\phi_k(w,\bar w)\}\ran_{str}^{conn}
\nonumber\\
&+\bigg(\frac{\mu_3}{4}\bigg)^2\frac{1}{2!}(\lan\{W_3W_3,
\phi_k(w,\bar w)\}\ran_{str}^{conn}
+2\lan W_3\phi_k(w,\bar w)W_3\ran_{str}^{conn})+\mathcal{O}(\mu_n^3)
\label{one-pt-fn-mu}
\end{align}
The $\{~,\}$ denotes an anticommutator. The operator ordering implies
the following: when $W_3$ appears on the left of $\phi_k(w,\bar w)$,
e.g., in $\lan W_3 \phi_k(w,\bar w) \ran$, the integration contour
\eq{contour} for $W_3$ on the strip lies above the point $(w,
\bar w)$; similarly when $W_3$ appears on the right of
$\phi_k(w,\bar w)$, e.g. in $\lan \phi_k(w,\bar w) W_3 \ran$, the
contour for $W_3$ is below the point $(w, \bar w)$.

The first, $\mu$-independent, term in the above expansion is already
calculated in \eq{one-pt-0}. 

\subsubsection{\label{sec:o-mu}$\mathcal{O}(\mu_n)$ Calculation}

As before, we find it convenient to use the conformal transformation
\eq{map}.  The correlator on the strip then reduces to that on the
UHP, as in the $\mu=0$ case before. For a holomorphic primary
field $\phi_k$, this gives
\begin{align}
\lan W_3\phi_k(w)\ran_{str}^{conn}=\frac{2\pi}{\b^2}
 z^{h_k}\left[i^3\int_{\Gamma_1}\!\!dz_1\ z_1^2
   \lan\wc_3(z_1)\phi_k(z)\ran_{\rm{UHP}}^{conn} 
   +(-i)^3\int_{\Gamma_1}\!\! d\bar{z}_1\ \bar{z}_1^2\lan\bar{\wc}_3(\bar
   z_1)\phi_k(z)\ran_{\rm{UHP}}^{conn}\right]
\label{str-o-mu}
\end{align}
where the operator ordering explained above implies that the contour
$\Gamma_1$ lies to the left of the point $(z, \bar z)$ on the
UHP. Now, in the analogous calculation \eq{cyl-o-mu}, the second
connected correlator on the complex plane vanished because of
factorization into one-point functions. Correlators on the UHP are,
however, related to those on the plane by the method of images (an
example of which we saw in \eq{images}).  In particular, $\bar{\wc}_3$
at the point $(z_1, \bar z_1)$ on the UHP becomes the holomorphic
operator $\wc_3^*= - \wc_3$ on the LHP at the point $(z_1', \zb_1')$
with $z_1'= \bar z_1$ \cite{Cardy:1984bb,DiFrancesco:1997nk}. The
contour $\Gamma_1$ gets mapped to its mirror image $\Gamma_1'$ on the
lower half plane.  With this, we get
\begin{align}
\lan W_3\phi_k(w)\ran_{str}^{conn}=\frac{2\pi}{\b^2}
 z^{h_k}\left[i^3\int_{\Gamma_1 + \Gamma_1'}\!\!dz_1\ z_1^2
   \lan\wc_3(z_1)\phi_k(z)\ran_{\mathbb{C}}^{conn} 
   \right]
\label{str-o-mu-C}
\end{align}
On the complex plane, the contour $\Gamma_1$ on the UHP can be
deformed to $\Gamma_1'$ on the LHP, hence the two contours simply
yield a factor of 2. In fact, combining with the other ordering, and
applying a similar reasoning, we get an overall factor of $4$. Thus,
combining with results from Section \ref{sec:one-pt-cyl}, we get, for
holomorphic primary fields
\begin{align}
-\frac{\mu_3}{4}
\lan \{W_3,\phi_k(w)\}\ran_{str}^{conn}=
- \mu_3 \lan W_3\phi_k(w)\ran_{cyl}^{conn}
\label{str-o-mu-4}
\end{align}
A similar statement is true for an antiholomorphic primary field.  

Let us turn now to primary fields $\phi_k(w, \bar w)$ with $h_k, \bar
h_k \ne 0$ (of the form $\phi_k(w,\bar w)= \varphi_k(w)\varphi_k(\bar
w)$, as discussed before in the context of \eq{images}).  In the
cylinder calculation in Section \ref{sec:one-pt-cyl} the
$\mu$-corrections for these vanished.  In the present case, they are
non-zero for operators of the form $\phi_k(w, \bar w)$=
$\varphi_k(w)\bar\varphi_k(\bar w)$, with $h_k = \bar h_k$ (as in
\eq{one-pt-prim-0}). After conformally transforming to the UHP, we
regard $\bar\varphi_k$ on the UHP as $\varphi_k^*$ at the image point on the
LHP (up to a constant). Combining with the arguments used for the
holomorphic operators, we eventually get
\begin{align}
&\frac{\lan \{W_3,\phi_k(w,\bar w)\} \ran_{str}^{conn}}
{\lan\phi_k(w,\bar w) \ran_{str}}=i^3 \frac{2\pi}{\beta^2} (z\zb)^h 
I_3(z,z'),
\nonumber\\
&I_3(z,z')\equiv 
\int_{\Gamma_1+\Gamma_1'+ \tilde{\Gamma}_1 +\tilde{\Gamma}'_1}\kern-30pt  
dz_1\ z_1^2 \lan\wc_3(z_1)\varphi_k(z)\varphi^*_k(z')\ran_{\mathbb{C}}^{conn}
/\lan\varphi_k(z)\varphi^*_k(z')\ran_{\mathbb{C}}^{conn}
\label{ratio-o-mu}
\end{align}

\begin{figure}[H]
\centering
\includegraphics[scale=.17]{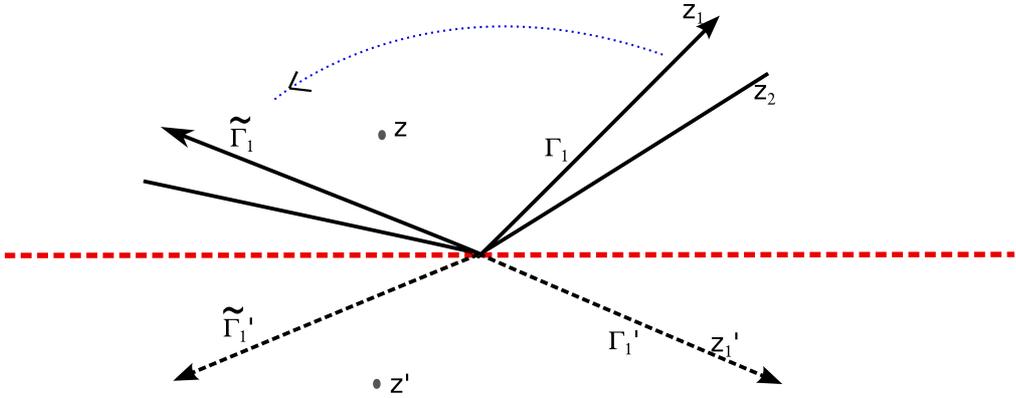}
\caption{\footnotesize Various contours needed to compute the $W_n$
  insertions in \eq{one-pt-fn-mu}. At late times, the insertion of
  each contour, irrespective of the position of the contour, amounts
  to insertion of a given factor linear in $t$.  This allows to resum
  arbitrary orders of arbitrary $W_n$-charge insertions, leading to
  the exponential time-dependence as in \eq{one-pt}. See Figure
  \ref{exponentiation-fig} for more.}
\label{contour-fig}
\end{figure}

The ratio of correlators inside the integral is given by
\begin{align}
\lan\wc_3(z_1)\varphi_k(z)\varphi^*_k(z')\ran_{\mathbb{C}}^{conn}/
\lan\varphi_k(z)\varphi^*_k(z')\ran_{\mathbb{C}}^{conn}= 
q_3\frac{(z-z')^3}{(z_1-z)^3(z_1-z')^3}
\end{align}
where $q_3$ is the $\wc_3$-charge of the field $\phi_k$.  Integrals of
the kind \eq{ratio-o-mu} are discussed in detail in Appendix
\ref{sec:integrals}. The final result (see \eq{i-3-long}) is that the
$O(\mu_3)$ correction, in the long time limit \eq{long-time}, is given
by (using that all four contours $\Gamma_1, \tilde \Gamma_1,
\Gamma_1', \tilde{\Gamma_1'}$ contribute equally, cancelling the
$1/4$ in $-\mu_3/4$)
\begin{align}
& \lan \phi_k(\s,t) \ran_{dyn} 
=  a_k e^{-2\pi \Delta_k t/\b} \left(1 - 
Q_{3,k} \tilde\mu_3 \ (\frac{2\pi t}{\b} + \hbox{constant})
+ O(\mu_3^2) \right) + ...,
\nonumber\\
& Q_{3,k} =i^3 2 q_{3,k} (2\pi),\; \tilde\mu_3= \frac{\mu_3}{\b^2}, 
\; \Delta_k = 2 h_k 
\label{o-mu-3}
\end{align}
Up to $O(\mu_3)$, it agrees with \eq{gamma-k}.

In case of a quasiprimary field $\phi_k(w,\bar w)$, it mixes, under
conformal transformation to the plane, with lower dimension
operators. The most relevant operator among these, which is of the
form $\varphi_k(z)\varphi_k(\bar z)$, is then to be used in
\eq{ratio-o-mu} for obtaining the dominant time-dependence; in that
case $\Delta_k, Q_{3,k}$ refer to this operator (rather than to the
original $\phi_k$).

For higher $W_n$ charges, the currents $\wc_n(w)$ are typically
quasiprimary, and hence they mix with lower order $\wc_m(z)$ under
conformal transformation to the UHP. Thus the $O(\mu_n)$ correction to
the dynamical one-point function $\lan \phi_k \ran_{dyn}$ is a linear
combination of terms of the form \eq{ratio-o-mu-n} (weighted by a set
of coefficients $a_{n,m}$, as in \eq{o-mu-n} below). Collecting all
this, the $O(\mu)$ correction with all chemical potentials is given by
\begin{align}
 \lan \phi_k(\s,t) \ran_{dyn} 
& =  a_k e^{-2\pi \Delta_k t/\b} \left(1 -
\sum_{n=3} Q_{n,k} \tilde\mu_n\ 
(\frac{2\pi t}{\b} + \hbox{constant})+ O(\mu^2) \right) + ...,
\nonumber\\
 \tilde\mu_n
& = \frac{\mu_n}{\b^{n-1}}, 
\; \Delta_k = h_k + \bar h_k= 2 h_k 
\nonumber\\
Q_{n,k} 
& =2 \sum_{m=0}^{\floor{n/2-1}} a_{n,m}  i^{n-2m} 
(2\pi)^{n-2m-2} q_{n-2m,k} 
\nonumber\\
& \kern80pt = i^n (2\pi)^{n-2} 2q_{n,k} + 
i^{n-2}(2\pi)^{n-4}a_{n,2}\, 2q_{n-2,k}
+ ... ,\; \label{o-mu-n}
\end{align}
Note that for $W_3$ deformations, the expression for $Q_3$ as in
\eq{o-mu-3} corresponds only to the first term in the above series
expression for $Q_n$. This is because the $\wc_3$ current is a primary
field and does not mix with any lower $\wc$ current under a conformal
transformation. From $n=4$ onwards, the additional terms in
$Q_{n,k}$'s represent the mixing of $\wc_n$ currents with $\wc_{n-2m}$
under conformal transformations.

\subsubsection{\label{sec:o-mu-all}Higher order $\mu$-corrections}

Let us first consider that $O(\mu_n^2)$ correction:
\begin{align}
\lan \phi_k(w,\bar w)\ran_{str}^{conn}|^{\mu_n}_2
\equiv \frac{(\mu_n/4)^2}{2!}(\lan\{W_nW_n,
\phi_k(w,\bar w)\}\ran_{str}^{conn}
 +2\lan W_n\phi_k(w,\bar w)W_n\ran_{str}^{conn})
\end{align}
Again, for holomorphic (or antiholomorphic) primary fields $\phi_k(w)$,
it is straightforward to generalize \eq{str-o-mu-4} to this order.
\begin{align}
\lan \phi_k(w)\ran_{str}^{conn}|^{\mu_n}_2
= \frac{\mu_n^2}{2!} \lan W_n W_n  \phi_k(w)\ran_{cyl}^{conn}
\label{str=cyl-2}
\end{align}
For a primary field of the form $\phi_k(w,\bar w)=
\varphi_k(w) \varphi_k(\bar w)$, proceeding as in the
previous subsection, we get
\begin{align}
&\lan \phi_k(w)\ran_{str}^{conn}|^{\mu_n}_2= \frac{1}{2!}
  \left(Q_{n,k} \tilde\mu_n t\ \frac{2\pi}{\b}\right)^2 +
\mu_n^2({\rm constant}\times t+ {\rm constant})+ ...
\label{str-o-mu-2}
\end{align}
The essential ingredient in this calculation is
\begin{align}
&I_{nm}(z,z'|\Gamma_1, \Gamma_2)\equiv 
\int_{\Gamma_1}\kern-3pt  
dz_1\ z_1^{n-1}\int_{\Gamma_2}\kern-3pt  
dz_2\ z_2^{m-1} f_{nm}(z_1,z_2,z,z'), \nonumber\\
&f_{nm}(z_1,z_2,z,z')= 
\frac{\lan\wc_n(z_1)\wc_m(z_2)\varphi_k(z)\varphi^*_k(z')
\ran_{\mathbb{C}}^{conn}
}{\lan\varphi_k(z)\varphi^*_k(z')\ran_{\mathbb{C}}^{conn}}
\label{ratio-o-mu-nn}
\end{align}
By repeating the strategy of \eq{residue-o-mu}, we get
\begin{align}
&\hbox{Coefficient of} ~ [\log(-z')- \log(-z)]^2~{\rm in}~
I_{nm}(z,z'|\Gamma_1, \Gamma_2)
\nonumber\\
&={\hbox{Residue}}_{z_1=z}\left[
{\hbox{Residue}}_{z_2=z}\left( \frac{\lan\wc_n(z_1)
\wc_m(z_2)\varphi_k(z)\varphi^*_k(z')\ran_{\mathbb{C}}^{conn}
}{\lan\varphi_k(z)\varphi^*_k(z')\ran_{\mathbb{C}}^{conn}}
\right)\right] =  q_{n,k}\, q_{m,k} 
\label{residue-o-mu-2}
\end{align}
where we have first used the $\wc_m(z_2)\varphi_k(z)$ OPE, and then
the $\wc_n(z_1)\varphi_k(z)$ OPE. In a manner similar to that in
Appendix \ref{sec:integrals}, we conclude the following structure
of $I_{nm}(z,z')$:
{\footnotesize
\begin{align}
I_{nm}(z,z'|\Gamma_1, \Gamma_2)=  
q_{n,k}\, q_{m,k} ([\log(-z')- \log(-z)]+ {\rm constant})
\times ([\log(-z')- \log(-z)]+ {\rm constant})
\end{align}
}
Note that at late times $t\gg \b$, $([\log(-z')- \log(-z)]$ $\to$
$2(2\pi t)/\b$ and dominates over the constant term (the precise sense
is that of \eq{leading-log}).  Similar to Appendix
\ref{sec:integrals}, the $4\times 4=16$ locations of the contour-pairs
$(\Gamma_1, \Gamma_2)$,$(\Gamma_1, \Gamma_2')$, $(\Gamma_1,
\tilde\Gamma_2)$,$(\Gamma_1, {\tilde\Gamma_2}')$, ...., all contribute
equally, therefore converting $(\mu_n/4)(\mu_m/4) \to$ $\mu_n
\mu_m$. Combining all these, we get \eq{str-o-mu-2}.  The charges
$q_n$ that are defined by the $\wc_n \varphi$ OPE and appear in
\eq{residue-o-mu-2}, get multiplied by some constants
\footnote{\label{ftnt:const}Each $\wc_n$ current comes with
a factor of  $\frac{i^n}{2\pi} \left(\frac{2\pi}{\b}\right)^{n-1}$,
as in \eq{contour-z}.} and shifted
by lower $\wc_{n-2k}$ charges to give the $Q_n$ in
\eq{residue-o-mu-2}, as in \eq{o-mu-n}.

\subsubsection*{\label{sec:exponentiation}Arbitrary 
orders and Exponentiation:} 

It is straightforward to generalize the above $O(\tilde\mu^2)$
calculation to higher orders in the perturbation in chemical
potentials.  Thus, at the order $\prod_{i=1}^r \mu_{n_i}$, there are
$r$ insertions of $\wc$-currents, leading to integrals of the form
\begin{align}
&I_{n_1 n_2... n_r}(z,z'|\Gamma_1, \Gamma_2,..., \Gamma_r)\equiv 
\int_{\Gamma_1}\kern-3pt  
dz_1\ z_1^{n_1-1}\int_{\Gamma_2}\kern-3pt  
dz_2\ z_2^{n_2-1}...\int_{\Gamma_r}\kern-3pt  
dz_2\ z_r^{n_r-1} f_{n_1 n_2...n_r}(z_1,z_2,...,z_r;z,z'), \nonumber\\
&f_{n_1 n_2...n_r}(z_1,z_2,...,z_r;z,z')=
\frac{\lan\wc_{n_1}(z_1)\wc_{n_2}(z_2)...
\wc_{n_r}(z_r)\varphi_k(z)\varphi^*_k(z')
\ran_{\mathbb{C}}^{conn}
}{\lan\varphi_k(z)\varphi^*_k(z')\ran_{\mathbb{C}}^{conn}}
\label{ratio-o-mu-nn...}
\end{align}
Again, repeating the strategy of \eq{residue-o-mu}, we get the
following leading ({\it viz.} $(\log)^r$) contribution
(see \eq{leading-log} for the definition of the leading-log 
contribution) 
{\footnotesize
\begin{align}
&\hbox{Coefficient of} ~ [\log(-z')- \log(-z)]^r~{\rm in}~
I_{n_1 n_2... n_r}(z,z'|\Gamma_1, \Gamma_2,..., \Gamma_r)
\nonumber\\
&={\hbox{Residue}}_{z_1=z}\left[...{\hbox{Residue}}_{z_{r-1}=z}\left\{
{\hbox{Residue}}_{z_r=z}\left( 
\frac{\lan\wc_{n_1}(z_1)...\wc_{n_{r-1}}(z_{r-1})\wc_{n_{r}}(z_{r})
\varphi_k(z)\varphi^*_k(z')\ran_{\mathbb{C}}^{conn}
}{\lan\varphi_k(z)\varphi^*_k(z')\ran_{\mathbb{C}}^{conn}}
\right)\right\}\right]
\nonumber\\
&=  q_{n_1,k}... q_{n_{r-1},k}\, q_{n_r,k} 
\label{residue-o-mu-r}
\end{align}}
where we have first used the $\wc_{n_r}(z_r)\varphi_k(z)$ OPE, then
$\wc_{n_{r-1}}(z_{r-1})\varphi_k(z)$ OPE, etc.  As in the
$O(\mu^2)$ calculation above, we obtain the following behaviour at
late times
{\footnotesize
\begin{align}
&  I_{n_1 n_2... n_r}(z,z'|\Gamma_1, \Gamma_2,..., \Gamma_r)
\nonumber\\
& =  q_{n_1,k}... q_{n_{r-1},k}\, q_{n_r,k} \,
  \underbrace{([\log(-z')- \log(-z)]+ {\rm constant})\times
 ...\times 
([\log(-z')- \log(-z)]+ {\rm constant})}_{r~\hbox{terms}}
\label{log-r}
\end{align}}
{\it The two equations above show that the leading
  log contribution to \eq{ratio-o-mu-nn...} from every contour
  integral of the $\wc_{n_i}$ current contributes the factor
  $q_{n_i}[\log(-z')- \log(-z)]$. This is the first basic ingredient
  for the exponentiation we are going to find}.  Furthermore, it is
easy to see that the leading log contribution is the same irrespective
of where each contour $\Gamma_i$ is placed (out of 4 possible choices,
e.g. $\Gamma_1, \Gamma'_1, \tilde \Gamma_1, \tilde \Gamma_1'$ in
Figure \ref{contour-fig}). As before we must combine the contribution
of all positions of the contours, which, therefore, amounts to
multiplying the result for \eq{ratio-o-mu-nn...} by $4^r$ which
converts the original coefficients coming from $\exp[-\sum_n \mu_n
  W_n/4]$ as follows
\[
\frac{\prod_{i=1}^r \mu_i/4}{r!} \to 
\frac{\prod_{i=1}^r \mu_i}{r!}.
\] 
{\it This is the second basic factor leading to the exponentiation.}
Combining all these, and incorporating some additional constants (see
footnote \ref{ftnt:const}) we get the following, leading, order $
(\mu_{n_1}...\mu_{n_r})$ contribution
\begin{align}
&{\lan \phi_k(w)\ran_{str}^{conn}}|^{\mu_{n_1}...\mu_{n_r}}_r= \frac{1}{r!}
 \prod_{i=1}^r \left(Q_{n_i,k} \tilde\mu_{n_i}\ 
\frac{2\pi}{\b}\right) + O(\mu^r t^{r-l})
\label{str-o-mu-nn...}
\end{align}
Once again, the constants $Q_n$ are related to the $q_n$ as in
\eq{o-mu-n}) in a manner similar to the $O(\tilde\mu)$ and the
$O(\tilde\mu^2)$ calculation above. We note that the leading log
contribution used in this paper can be isolated by considering a
scaling
\begin{align}
\tilde\mu_n \to 0, \tilde t\equiv
\frac{t}{\b} \to \infty, ~\hbox{such that}~ 
\tilde\mu_n \tilde t= \hbox{constant}. 
\label{leading-log}
\end{align}
The second term in \eq{str-o-mu-nn...}, or for that matter, in
\eq{str-o-mu-2}, is subleading at large times in the sense of this
scaling.

Using the above results, we now have, for primary fields of the form
$\phi_k(w,\bar w)=\varphi_k(w)\varphi_k(\bar w)$
\begin{align}
\lan \phi_k(w,\bar w) \ran_{str} 
= a_k e^{-\frac {2\pi \Delta_k t}{\b}}& \left[1 - \sum_n  \tilde\mu_n\  Q_{n,k}
(\frac{2\pi t}{\b} + {\rm const}) \right. 
\nonumber\\
&\left. + \frac1{2!} 
\sum_{n,m} \tilde \m_n \tilde\mu_m\  Q_{n,k}(
\frac{2\pi t}{\b} + {\rm const})
\  Q_{m,k}(\frac{2\pi t}{\b}
 + {\rm const}) \right. + ...
\nonumber\\
&\left. + \frac1{r!} \kern-2pt
\sum_{\{n_i\}}\kern-2pt\prod_{i=1}^r 
\tilde \m_{n_i} Q_{n_i,k}\left(
\underbrace{(\frac{2\pi t}{\b} + 
{\rm const})...(\frac{2\pi t}{\b} + {\rm const})}_{r\, {\rm terms}}\right)  
+ ...\right]
\nonumber\\
& = a_k e^{-2\pi  t/\b \left( \Delta_k +\sum_n \tilde \m_nQ_{n,k} +
O({\tilde\mu}^2) \right)} =  a_k e^{-\g_k t}
\label{final-one-pt}
\end{align}
where we have absorbed some constant factors in $a_k$.  $\g_k$ is
given by \eq{gamma-k}; $Q_{n,k}$ are the shifted $W_n$ charges of
$\phi_k$ as defined in \eq{o-mu-n}. The proof of the above equation
for general quasiprimary operators $\phi_k(w,\bar w)$ works out much
the same way as in case of the $O(\mu)$ terms, as discussed in Section
\ref{sec:o-mu}. {\it We emphasize that it is only the leading
  contributions at large times which we have proved here to
  exponentiate.} Thus, we do not claim that the constant terms marked
``const'' in the above equation are all the same. As we have remarked
before, the leading contributions can be isolated using the scaling
mentioned in \eq{leading-log}. 

The schematics of the above calculation is explained in the Figure
\ref{exponentiation-fig}.

\begin{figure}[H]
\centering
\includegraphics[scale=.3]{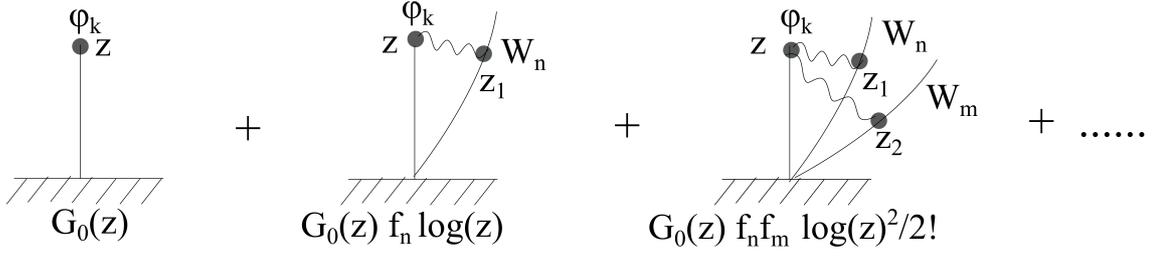}
\caption{\footnotesize The schematics of the calculation of the
  one-point function. The first term represents the zero-order
  boundary Green's function \eq{one-pt-prim-0} without chemical
  potentials (the shading indicates the boundary of the upper half
  plane). The second term represents the $O(\mu_n)$ correction, which
  involves one insertion of a $W_n$-charge (which is an integral over
  the $z_1$-contour. As explained in the text, at long times, this
  insertion amounts to multiplying the zero order term by a term of
  the form $f_n\log(z)$, where $f_n$ is described in \eq{o-mu-n}. The
  third term represents insertion of two such $W$-charges; as we
  explained in the text (see \eq{str-o-mu-2} and below), each
  insertion again amounts to multiplying by the factor mentioned
  above, along with a factor of $\frac1{2!}$. The pattern continues,
  to ensure an exponentiation to $G_0(z)\, z^{\sum_n\kern-2pt f_n}$,
  as in \eq{final-one-pt}. Since at long times $G_0(z) \sim
  e^{-\g^{(0)}_k t}$ (see \eq{one-pt-prim-0}), and $z \sim e^{-2\pi
    t/\b}$, adding the chemical potentials amount to a shift of the
  exponent $\g^{(0)}_k \to \g_k$ as in \eq{one-pt}.}
\label{exponentiation-fig}
\end{figure}

\section{\label{sec:i-t}Calculation of $I(t)$}

Let us rewrite the expression for the thermalization
function $I(t)$ \eq{I-t} in the form 
\begin{align}
&I(t) = Z_{sc}/
\sqrt{Z_{ss} Z_{cc}}  = \hat Z_{sc}/
\sqrt{\hat Z_{ss} \hat Z_{cc}},
\nonumber\\
& Z_{sc} \equiv \Tr(\rho_{dyn,A}(t)  \rho_{eqm,A}(\b, \mu)),\,
\hat Z_{sc} =  Z_{sc}/(Z_s  Z_c) 
\nonumber\\
&Z_{ss} \equiv \Tr(\rho_{dyn,A}(t) \rho_{dyn,A}(t)),\,
\hat Z_{ss} =  Z_{ss}/Z_s^2,
\nonumber\\
& Z_{cc} \equiv \Tr(\rho_{eqm,A}(\b, \mu) \rho_{eqm,A}(\b, \mu) ),\,
\hat Z_{cc} =  Z_{cc}/Z_c^2,
\nonumber\\
& Z_s= \Tr(\rho_{dyn}(t))= \lan \psi_0 | \psi_0 \ran, \;
Z_c= \Tr(\rho_{\b,\mu})
\label{I-t-zs}
\end{align}
In Appendix \ref{sec:short} we explain how to compute $I(t)$ using the
short interval expansion, valid when the length of the interval $l$ is
small compared with the other time scales $\b$ and $t$ in the problem.
We reproduce the main formula \eq{short-interval} for our purpose,
where we explicitly denote the dependencies on the length $l$
of the interval, the inverse temperature $\b$ and the chemical
potentials $\mu$ (the dependence on $\b$ on the RHS is implicit;
the one-point functions depend on both $\b$ and $\mu$--- see
Section \ref{sec:one-pt}). 
\begin{align}
& \hat Z_{sc}(l,\b,\mu) =\sum_{k_1, k_2} 
C_{k_1,k_2}(l) \lan \phi_{k_1}(w_1,
\wb_1) \ran_{str}^\mu \lan
  \phi_{k_2}(w_2, \wb_2) \ran_{cyl}^\mu, \nonumber\\
& \hat Z_{ss}(l,\b,\mu) =\sum_{k_1, k_2}
  C_{k_1,k_2}(l) \lan \phi_{k_1}(w_1,\wb_1) \ran_{str}^\mu \lan \phi_{k_2}(w_2, \wb_2)
  \ran_{str}^\mu,\, \nonumber\\
&\hat Z_{cc}(l,\b,\mu) =\sum_{k_1, k_2} C_{k_1,k_2}(l) \lan
  \phi_{k_1}(w_1,\wb_1) \ran_{cyl}^\mu 
\lan \phi_{k_2}(w_2, \wb_2) \ran_{cyl}^\mu
\label{short-interval-mu}
\end{align}
It is understood, for the logic of the short interval expansion
to go through, that all contours which represent insertion
of the $W$-charges (see Fig \ref{contour-fig}) are drawn outside
of the small disc-like region of both sheets of Fig \ref{tonni-fig}.
 
\subsection{{\label{therm}}Proof of thermalization}

Using the short-interval expansion above, and the long time behaviour
of one-point functions from Section \ref{sec:one-pt}), it is easy to
prove that the system thermalizes in the sense of \eq{I-infinity} or
\eq{rhoA-infinity}.

To prove this, note that it is only the holomorphic (or
antiholomorphic) fields $\phi_k$ which possibly have non-zero
expectation values in the long time limit \eq{long-time}.  For these
fields, the one-point functions on the cylinder and on the strip agree
(see \eq{one-pt-0}, \eq{str-o-mu-4}, \eq{str=cyl-2} ). By virtue of
\eq{short-interval-mu}, we therefore have in the long time limit
$Z_{sc}= Z_{ss}= Z_{cc}$. Hence using the expression \eq{I-t-zs} for
the thermalization function we get $I(t\to \infty)=1$ which proves
\eq{I-infinity} and consequently \eq{rhoA-infinity}.

The above-mentioned equality of one-point functions between the strip
and cylinder geometries for holomorphic (or antiholomorphic) fields
imply the same for the conserved $\wc_n$- (or $\bar\wc_n$)- currents.
This, therefore, proves that
\begin{align}
\lan \psi(t) |  W_n  | \psi(t) \ran =  \Tr(W_n \rho_{eqm})
\label{charge-match}
\end{align}
Note that in proving this, we have used the correspondence 
\eq{eq-ensemble} between the parameters of the initial state
and the putative equilibrium state. The above equation, therefore,
proves the correspondence \eq{eq-ensemble}.

\subsection{\label{sec:gamma-mu0}Thermalization rate}

To evaluate the rate of approach of $I(t)$ to its asymptotic value 1,
we organize the terms in $\hat Z_{sc}, \hat Z_{ss}, \hat Z_{cc}$ as
follows
\begin{align}
& \hat Z_{sc} = C_{0,0}(1+ S_1^{sc}),\;
S_1^{sc}=  \sum_a \hat C_{a,0}(\lan \phi_a \ran^\mu_{str} + 
\lan \phi_a \ran_{cyl}^\mu) + \sum_{ab}
\hat C_{a,b} \lan \phi_a \ran_{str}^\mu
\lan \phi_b \ran_{cyl}^\mu
\nonumber\\
& \hat Z_{ss} = C_{0,0}(1+ S_1^{ss}+ S_2^{ss}),\;
S_1^{ss} = 2 \sum_a \hat C_{a,0}
\lan \phi_a \ran_{str}^\mu
+ \sum_{ab} \hat C_{a,b} \lan \phi_a \ran_{str}^\mu
\lan \phi_b \ran_{str}^\mu,\;
S_2^{ss}=  \sum_{k} \hat C_{k,k} 
(\lan \phi_k \ran_{str}^\mu)^2  
\nonumber\\
& \hat Z_{cc} = C_{0,0}(1+ S^{cc}_1),\;  S^{cc}_1=
2 \sum_a \hat C_{a,0}\lan \phi_a \ran_{cyl}^\mu
+ \sum_{ab} \hat C_{a,b} \lan \phi_a \ran_{cyl}^\mu
\lan \phi_b \ran_{cyl}^\mu
\label{n-d1-d2-form}
\end{align}
where $a,b,...$ denote descendents of the identity operator,
$k$ labels other primaries (than the identity) and their descendents.
$\hat C \equiv C/C_{0,0}$.  

\subsubsection{$\mu=0$}

Let us first consider the case of zero chemical potentials.  Using the
results in Sections \ref{sec:one-pt}, and Appendices \ref{app:one-pt}
and \ref{app:ck1k2}, we get
\begin{align}
& S_1^{sc}= - a_T \tilde l^2 \left(1+ O(\tilde l)^2\right)
+ a_{T\bT}  \tilde l^4 e^{-8\pi t/\b}\left(1+ O(\tilde l)^2\right)
+ O(e^{-8\pi\tilde t})
\nonumber\\
& S_1^{ss}= - a_T \tilde l^2 \left(1+ O(\tilde l)^2\right)
+ 2 a_{T\bT}  \tilde l^4 e^{-8\pi\tilde t}\left(1+ O(\tilde l)^2\right)
+ O(e^{-8\pi\tilde t})
\nonumber\\
& S_2^{ss}= \sum_k\left[ a_k \tilde l^{4h_k}e^{-8\pi h_k t/\b}
\left(1+ O(\tilde l)^2\right) +  
O(e^{-12\pi h_k\tilde t}) \right]
\nonumber\\
& S_1^{cc}= - a_T \tilde l^2 \left(1+ O(\tilde l)^2\right)
\nonumber\\
& a_T= \frac{c \pi^2}{24},\;  a_{T\bT} = \frac{A_{T\bT} \pi^4}{8 c}\;
a_k=  \frac{A_k^2}{n_k} \left(\frac{\pi}{2}\right)^{4h_k}
\end{align}
To this order, it is easy to see that the contribution to $I(t)$ from
descendents of identity, demarcated by $a_T, a_{T\bT}$, vanishes. The
leading contribution to $I(t)$, demarcated by $a_k$, occurs only in
$\hat Z_{ss}$ and comes from
$\left(\lan\phi_m(z,\zb)\ran_{str}\right)^2$ for which $h_k$ is the
minimum ($= h_m$) (this could be a field which appears after a
conformal transformation of the original quasiprimary field). The
time-dependence shown of $S_2^{ss}$ comes from
\eq{one-pt-prim-0}. Using this, we get
\begin{align}
I(t) &= 
1 -  \alpha \exp[-2\g_m^{(0)} t] + ...,\;  \g_m^{(0)}= 
2 \pi \Delta_m/\b
\label{I-t-mu=0}
\end{align}
This is of the form \eq{I-asymp} for $\mu=0$, with 
\begin{align}
\alpha \equiv \frac{A_m^2}{n_m} \left(\frac{\pi}{2}
\right)^{4h_m} 
\,(\tilde l)^{4 h_m}
\left(1+ O(\tilde l)^2\right)
\label{phi-sq}
\end{align}
The discarded terms in \eq{I-t-mu=0} are faster transients. This
proves \eq{I-asymp} for zero chemical potential. This result has
already appeared in
\cite{Cardy:2014rqa}.\footnote{\label{ftnt:cardy}Our exponent differs
  from Cardy's value by a factor of 2.}

\subsubsection{$\mu \ne 0$}

The generalization of the above result to the case of non-zero
chemical potentials is straightforward. Once again, the dominant
time-dependence arises from
$\left(\lan\phi_m(z,\zb)\ran_{str}^\mu\right)^2$ in the $S_2^{ss}$ or
$\hat Z_{ss}$.  The time-dependence \eq{I-asymp} follows by
using \eq{final-one-pt} in $S_2^{ss}$.

\subsection{\label{sec:q-hat}Properties of $\hat Q$}

From the asymptotic behaviour \eq{I-asymp} of the thermalization
function we indicated the asymptotic behaviour \eq{q-hat} of the
dynamical reduced density matrix $\hat \rho_{dyn}(t)$. By using the
long time behaviour of the one-point functions \eq{one-pt}, we can
easily deduce the following dominant behaviour of overlaps of $\hat Q$
with various quasiprimary fields at late times
\[
\Tr(\hat Q \phi_k(t)) \propto e^{-(\g_k - \g_m)t},
\; \Tr(\hat Q \phi_m(t))  \to {\rm constant}.
\]

\section{\label{sec:qnm-cft}Decay of perturbations of a thermal state}

We found in the previous sections that the long time behaviour of the
reduced density matrix $\rho_{dyn,A}(t)$ resembles that of a thermal
ensemble plus a small deformation which decays exponentially. We will
find in the next section that the thermal ensemble (or more accurately
the generalized Gibbs ensemble) corresponds to a (higher spin) black
hole geometry in the bulk. The small perturbation of the equilibrium
ensemble is thus expected to correspond to a small deformation of the
black hole geometry. Consequently, the exponential decay of the
deformation in the CFT should correspond to a `ringing-down' or a
quasinormal mode in the bulk.

We will address the above issue in the next section which deals with
bulk geometry. However, in order to make the correspondence of the
above paragraph more precise, in this section we will directly present
a CFT computation of the decay of a perturbation to a thermal state.
Note that this computation is, in principle, different from the
exponential decay of the one-point function in the quenched state,
\eq{one-pt}. However, what we will find is that the long time
behaviour \eq{one-pt} of an operator $\phi_k(0,t)$ in the quenched
state is the same as that of its two-point function
\eq{two-pt-thermal} in the thermal state \eq{ensemble} (with chemical
potentials). The latter measures the thermal decay of a perturbation
and is more directly related to a black hole quasinormal mode.
Throughout this section, we will assume that the conformal dimensions
of $\phi_k$ satisfy $h_k = \bar h_k$. 

We define the thermal two-point function as \footnote{We use the same
notations as in \cite{Festuccia:2006sa}.}
\begin{align}
G_+(t,0;\b,\mu)
\equiv \frac1{Z} \Tr(\phi_k(0,t) \phi_k(0,0) e^{-\b H -\sum_n \mu_n W_n})
\label{two-pt-thermal}
\end{align}
By the techniques developed in the earlier sections, a computation of
this quantity amounts to calculating the following correlator
on the plane
\begin{align}
\lan \phi_k(z, \bar z) \phi_k(y, \bar y) e^{-\sum_n \mu_n W_n}
\ran, \kern10pt   
z= i e^{- 2\pi t/\b}, \zb = - i e^{2\pi t/\b}, 
y= i, \bar y = -i 
\label{auto-plane}
\end{align}
where the $\mu_n$-deformations are understood as an infinite series of
contours as in the previous section. 

For $\mu=0$, the above two-point function is standard. Including
the Jacobian of transformation, we get 
\begin{align}
G_+(t,0;\b,0)= (\frac{2\pi}\b)^{4 h_k} \left[(i e^{- 2\pi t/\b} - i)
  (- i e^{2\pi t/\b} + i)\right]^{-2 h_k} \xrightarrow{t\to \infty}
\hbox{const}~e^{-2\pi t \Delta_k/\b}, 
\label{auto}
\end{align}
which clearly matches the long time behaviour of the
one-point function \eq{one-pt} in the quenched state for $\mu=0$.
Here $\Delta_k= 2 h_k.$

In the above, we considered the thermal Green's function for two
points which are both at the same spatial point $\s=0$. It is easy to
compute the Green's function when the two points are spatially
separated by a distance $l$, say with $\s_1=l$ and $\s_2=0$. We get
\begin{align}
G_+(t,l;\b,0)
& \equiv \frac1{Z} \Tr(\phi_k(l,t) \phi_k(0,0)e^{-\b H})
 =\left[\frac{2\pi}\b e^{\pi l /\b} \right]^{4 h_k}
\kern-10pt \left((i e^{2\pi(l- t)/\b} - i)
  (- i e^{2\pi(l+ t)/\b} + i)\right)^{-2 h_k} 
\nonumber\\
&\kern60pt \xrightarrow{t, l \gg \b}
\left\{\begin{array}{ l l}
\hbox{const}~e^{-2\pi t \Delta_k/\b}, & (t-l) \gg \b
\\
\hbox{const}~e^{-2\pi l \Delta_k/\b}, &  (l-t) \gg \b 
\end{array}\right. 
\label{thermal-decay}
\end{align}
The coordinates of the two points, in the notation of \eq{auto-plane}
are modified here to $z= i e^{2\pi(l- t)/\b}, \zb = - i e^{2\pi(l+
  t)/\b},$ $y= i, \bar y = -i $. The prefactor with the square
bracket comes from the Jacobian of the transformation from the
cylinder to the plane.
The behaviour of the Green's function is shown in Figure
\ref{thermal-fig}. It is important to note that the exponential decay,
found in \eq{one-pt} shows up only for time scales $t\gg l$.

\begin{figure}[H]
\centering
\includegraphics[width=275pt, height=150pt]{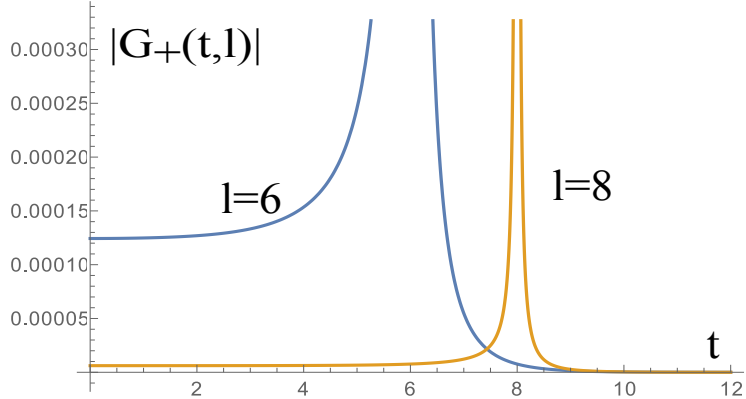}
\caption{\footnotesize Plots of the thermal Green's function
  $G_+(t,l;\b,0)$ for $\b=2\pi$, $\Delta_k= 1.5$.  The curve on the
  left (blue) is for $l= 6$, and the curve on the right (orange) is
  for $l= 8$. Note that the exponential decay in time occurs for times
  larger than $l$.}
\label{thermal-fig}
\end{figure}

The effect of turning on the chemical potentials can be dealt with as
in the previous sections. At $O(\mu_n)$, we will have, as before,
a holomorphic contribution and an antiholomorphic contribution.
The former is proportional to 
\begin{align}
\lan \phi_k(\bar z) \phi_k(\bar y) \ran \times
\int_\Gamma dz_1 z_1^{n-1} \lan \wc_n(z_1) \phi_k(z) \phi_k(y) \ran
\end{align}
As we see, the structure of the integral is the same as in the
previous section. As before, logarithmic terms appear in the above
integrals which give the leading, linear, $t$-dependence. Similar
remarks also apply to the antiholomorphic contour.  Since the
calculations are very similar to those in the previous two sections,
we do not provide all details. By resumming the series
over the infinite number of contours, we find in a straightforward
fashion that 
\begin{align}
G_+(t,0;\b,\mu)
\xrightarrow{t\to \infty} G_+(0,0;\b,0) b(\mu) e^{-\g_k t}
\label{two-pt-thermal-asym}
\end{align}
where $b(\mu)$ is time-independent, and is of the form $b(\mu)= 1 +
O(\mu)$. This long time decay is the same as that of the one-point
function \eq{one-pt} in the quenched state, as claimed above.  For
points separated by a distance $l$, the above exponential decay
shows up for $t \gg l$, as in  \eq{thermal-decay}. 

In the above, we have discussed the two-point function in real
space. It is straightforward to convert the result \eq{thermal-decay}
without chemical potentials to Fourier space, which develops poles at
\begin{align}
\omega_{k,m}|_{_{\mu=0}} = -i \frac{2\pi}{\b} (\Delta_k + 2m), \, m=0,1,2,...
\label{cft-qnm-beta}
\end{align}
Our results in \eq{one-pt} can be interpreted as a shift, caused
by the presence of the chemical potentials $\mu_n$, of the dominant
pole $\om_{k,0}|_{_{\mu=0}}$ to 
\begin{align}
\omega_{k,0}= -i \frac{2\pi}{\b} (\Delta_k +
\sum_n \tilde \mu_n Q_{n,k})
= -i \g_k, 
\label{cft-qnm-beta-mu}
\end{align}
where the notation is the same as that of \eq{one-pt}. In this paper
we will not address the question of the shift of the subdominant poles
$\om_{k,m}$ (for $m=1,2,...$) due to chemical potentials (the current
status of these can be found in \cite{Cabo-Bizet:2014wpa,
  Gaberdiel:2013jca, Beccaria:2013yca}).
 
Two-point functions of the kind \eq{two-pt-thermal}, for a single
chemical potential $\mu_3$, and up to order $\mu_3^2$, have appeared
earlier in \cite{Gaberdiel:2013jca} (calculations up to $O(\mu_3^5)$
have appeared in \cite{Beccaria:2013yca}). What we find in our paper
is that at large times, the perturbation series in $\mu_n$, up to all
orders in all chemical potentials, can be resummed, to yield the
leading correction to the thermalization rate in the presence of
chemical potentials.

At a technical level, the one-point function in the quenched state
corresponds to a one-point function in a geometry with a boundary, and
for operators considered here, these turn into a two-point function on
the plane, by virtue of the method of images. The thermal decay
naturally involves a two-point function on the
plane \footnote{Actually the thermal calculation involves a product of
  two such factors, one holomorphic and the other antiholomorphic, but
  one of the factors just gives an overall constant and only one
  factor leads to the important time-dependence.} and agrees with the
above two-point function at late times. 

\section{\label{sec:qnm}Holography and higher spin black holes}

\paragraph{Zero chemical potential:}
As remarked in the Introduction, a global quantum quench described by
an initial state of the form \eq{psi-0}, for large central charges and
zero chemical potentials, has been shown in \cite{Maldacena:2001kr,
  Hartman:2013qma,Caputa:2013eka} to be dual to one half of the
eternal BTZ (black string) geometry, whose boundary represents an
end-of-the-world brane. 


In an independent development, it was found in
\cite{Birmingham:2001pj} that the quasinormal mode of a scalar field
$\Phi_k(\s,t,z)$ of mass $m$ in a BTZ background (dual to a CFT
operator $\phi_k$ of dimension $\Delta_k$ $\equiv $ $ 1
+ \sqrt{1+m^2}$) is of the form
$\exp[-2\pi \Delta t/\b]$ at large times. 
This time-dependence agrees with the CFT exponent in
\eq{thermal-decay} exactly. This shows that the exponential decay of a
CFT perturbation to a thermal state corresponds to the decay of the
corresponding scalar field in the bulk geometry. This result has been
extended to higher spin fields in the BTZ background in
\cite{Datta:2011za}.

\paragraph{Non-zero chemical potentials:} In case the CFT
has additional conserved charges, in particular if it has a
representation of a $W_\infty$ algebra (and consequently the hs$(\l)$
algebra \cite{Gaberdiel:2010pz}), then the bulk dual corresponding to
those conserved charges have been conjectured to be the conserved
higher spin charges of higher spin gravity. In particular,
\cite{Gutperle:2011kf,Kraus:2011ds} have shown that if one interprets
the grand canonical ensemble \eq{eq-ensemble} (more generally, the
GGE) in the framework of an hs$(\l)$ representation, then the bulk
dual corresponds to a higher spin black hole.

Thus, we would like to conjecture that the bulk dual of the quantum
quench with chemical potentials, would correspond to a gravitational
collapse to a higher spin black hole.

As an important consistency check, by analogy with the case with zero
potential, in the present case too, the leading quasinormal mode (QNM)
of a scalar field $\Phi_k(\s,t,z)$ should have a time-dependence given
by \eq{two-pt-thermal-asym}. Following the results in
\cite{Cabo-Bizet:2014wpa} (see also
\cite{Gaberdiel:2013jca,Beccaria:2013yca,GM:2015Progress1}) 
\footnote{We wish to thank Alejandro Cabo-Bizet and Viktor
  Giraldo-Rivera for informing us that the difference between
  equation \eq{bulk-qnm-beta-mu} above and the corresponding equation (4.2)
  in a previous version of their paper \cite{Cabo-Bizet:2014wpa} was
  due to a typo, which has now been corrected in the new version of
  their paper.} we find that at late times $t\gg \b$ the QNM for the
hs$(\l)$ scalar field $\Phi_+$ behaves, up to $O(\mu_3)$, as $e^{-
  i\om_{_{k,0}} t}$, where
\begin{align}
 \om_{k,0}= -i \frac{2\pi}{\b} \left(1+\l + \tilde \mu_3
\frac13(1+\l) (2 + \l)\right)
\label{bulk-qnm-beta-mu}
\end{align} 
where the index $k$ here refers to the operator $\phi_k$ dual to the
scalar field $\Phi_+$.  Noting that for this operator we have
$\Delta_k= 1+ \l$, and $Q_{3,k} = \frac13(1+\l) (2 + \l)$
\cite{Gaberdiel:2013jca,Beccaria:2013yca}, we see that the QNM
frequency $\om_{k,0}$ agrees, to the relevant order, with the pole
\eq{cft-qnm-beta-mu} of the thermal 2-point function which, in turn,
is related to the thermalization exponent by the relation
$\om_{k,0}=-i \gamma_k$, with $\gamma_k$ given in \eq{one-pt}.

\section{\label{sec:discuss}Discussion}

In this paper, 2D conformal field theories were considered with
additional conserved charges besides the energy. We probed
non-equilibrium physics starting from global quenches described by
conformal boundary states modified by multiple UV cut-off parameters
\eq{psi-ep-0}. It was found that local observables in such a state
thermalize to an equilibrium described by a grand canonical ensemble
\eq{eq-ensemble} with temperature and chemical potentials related to
the cut-off parameters. We computed the thermalization rate for
various observables, including the reduced density matrix for an
interval. It was found that the same rate appears also in the long
time decay of two-point functions in equilibrium (see \eq{one-pt}
and \eq{thermal-decay-mu}). In the context where
the number of conserved charges is infinite, and they are identified
with commuting $W_\infty$ charges, the equilibrium ensemble (a generalized
Gibbs ensemble, GGE)
corresponds to a higher spin black hole
\cite{Gutperle:2011kf,Kraus:2011ds}.  We found that the thermalization
rate found above agrees with the leading quasinormal frequency of the
higher spin black hole; this constitutes an additional, dynamical,
evidence for the holographic correspondence between the global
quenches in this paper and the evolution into the higher spin black
hole.

One of the main technical advances made in this paper is the
resummation of leading-log terms at large times, presented in Section
\ref{sec:o-mu-all}, which leads to exponentiation of the perturbation
series, leading to the thermalization rate, presented in \eq{one-pt},
\eq{final-one-pt}, as a function of chemical potentials. This allows
us to also compute the effect of chemical potentials on the relaxation times
of thermal Green's functions. Another technical advance 
consists of the computation of the long-time reduced density matrix
\eq{I-asymp}, using a short-interval expansion, 
which allows us to prove thermalization of an
arbitrary string of local observables.

One might wonder whether the results presented in this paper are tied
to the use of translationally invariant quenched states such as
\eq{psi-ep-0}, whose energy density and various charge densities are
uniform. We will address the question of inhomogeneous quench in a
forthcoming paper \cite{GM:2015Progress2}, both in the CFT and in the
holographic dual, using the methods of \cite{Mandal:2014wfa} where we
create an inhomogeneous energy density by applying conformal
transformations. It turns out \cite{GM:2015Progress2} that if the
initial state has inhomogeneities in a compact domain and has uniform
energy densities outside, local observables again thermalize
asymptotically with exponents governed by the uniform densities. Other
important issues involve local quenches (see,
e.g. \cite{Calabrese:2007-local, Nozaki:2013wia}), and compact spatial
dimensions. The issue of thermalization when
space is compact is quite subtle. It has been shown in \cite{Cardy:2014rqa} that at
large times one can have the phenomenon of revival (observables
effectively returning to their initial values). The dynamical
entanglement entropy for a quantum quench in a space with boundaries
is an interesting, related, issue; we hope to come back to this in a
forthcoming publication \cite{GM:2015Progress3}.

\subsection*{Acknowledgement}

We would like to thank Pallab Basu, Justin David, Deepak Dhar, Oleg
Evnin, Rajesh Gopakumar, Shiraz Minwalla, Pranjal Nayak, Arunabha Saha
and Tomonori Ugajin for discussions and Somyadip Thakur for
discussions and collaboration in the forthcoming paper
\cite{GM:2015Progress1}, partial results from which are presented
here. We would also like to thank Juan Pedraza for drawing our
attention to Ref. \cite{Caceres:2014pda} and Alejandro
Cabo-Bizet and Viktor Giraldo-Rivera for a useful correspondence
regarding Ref. \cite{Cabo-Bizet:2014wpa}.

\appendix

\section{\label{app:one-pt}Some details on
one-point functions}

Here we collect some additional helpful material on the
one-point functions discussed in this paper.

\subsection{\label{app:one-pt-list}A few explicit one-point
functions with zero chemical potentials}

\ni\underbar{Case $k=$ descendent of identity}: In this case,
$\phi_k(w,\wb)$ is of the form $T, \bT$, or $\nor{T\bT}$ or some
descendents thereof. Under a conformal transformation \eq{map}, these
operators pick up a c-number term in addition to a term proportional
to the corresponding operator on the plane/UHP. We will give
some examples to illustrate the calculation\\
\ni 1. {\it cylinder:}
In this case 
\begin{align}
&\lan T(w)\ran_{cyl} = \lan \left(-\frac{c\pi^2}{6 \b^2} -\frac{4\pi^2}
{\b^2} z^2 T(z)\right) \ran_{UHP}= -\frac{c\pi^2}{6 \b^2}
\nonumber\\
& \lan \nor{T\bT}(w,\wb) \ran_{cyl} = 
\lan \left([-\frac{c\pi^2}{6 \b^2} -\frac{4\pi^2}
{\b^2} z^2 T(z)][-\frac{c\pi^2}{6 \b^2} -\frac{4\pi^2}
{\b^2} \zb^2 \bT(\zb)] \right) \ran_{UHP}
= (\frac{c\pi^2}{6 \b^2})^2 
\label{tt-cyl}
\end{align}
\ni 2. {\it strip:} In this case
\begin{align}
\lan T(w)\ran_{str} &= \lan \left(-\frac{c\pi^2}{6 \b^2} -\frac{4\pi^2}
{\b^2} z^2 T(z)\right) \ran_{UHP}= -\frac{c\pi^2}{6 \b^2}= \lan T(w)\ran_{cyl}
\nonumber\\
\lan \nor{T\bT}(w,\wb) \ran_{str} &= 
\lan \left([-\frac{c\pi^2}{6 \b^2} -\frac{4\pi^2}
{\b^2} z^2 T(z)][-\frac{c\pi^2}{6 \b^2} -\frac{4\pi^2}
{\b^2} \zb^2 \bT(\zb)] \right) \ran_{UHP}
\nonumber\\
&= (\frac{c\pi^2}{6 \b^2})^2 + A_{T\bT}(z-\zb)^{-4}
=   (\frac{c\pi^2}{6 \b^2})^2 +  a_{T\bT} e^{-8\pi t/\b}+ ...
\label{tt-str}
\end{align}
where $A_{T\bT}$, $a_{T\bT}$ are
constants as  in \eq{one-pt-prim-0} and \eq{images}.\\ 
\ni\underbar{Case $k=$ descendent of other primaries}: In this case,\\
\ni 1. {\it cylinder:} The one-point function vanishes as
in the case of primaries.
\\
\ni 2. {\it strip:} The one-point function can be related to
one-point function of primaries which is dealt with above.

\subsection{\label{sec:integrals}Some details on $O(\mu_n)$
correction to the one-point function}

In this section we will consider the following integrals which
arise in connection with $O(\mu_n)$ correction to the one-point
function $\lan \phi(\s,t) \ran_{dyn}$:
\begin{align}
&I_n(z,z'|\Gamma_1)\equiv 
\int_{\Gamma_1}\kern-3pt  
dz_1\ z_1^{n-1} f_n(z_1,z,z'),\;\; g_n(z_1,z,z') \equiv \int 
dz_1\ z_1^{n-1} f_n(z_1,z,z') \nonumber\\
&f_n(z_1,z,z')= \frac{\lan\wc_n(z_1)\varphi_k(z)\varphi^*_k(z')
\ran_{\mathbb{C}}^{conn}
}{\lan\varphi_k(z)\varphi^*_k(z')\ran_{\mathbb{C}}^{conn}}
=q_{n,k}\frac{(z-z')^n}{(z_1-z)^n(z_1-z')^n}
\label{ratio-o-mu-n}
\end{align}
The second integral on the first line is an indefinite integral.
The integrals above can be explicitly computed. E.g.
\begin{align}
& g_3(z_1,z,z')
=q_{3,k}[R_3(z,z')(\log(z_1-z)-\log(z_1-z'))-\frac{z^2}{2(z_1-z)^2}+\frac{z'^2}{2(z_1-z')^2}\nonumber\\
&\kern200pt+\frac{z'(2z+z')}{(z-z')(z_1-z')}+\frac{z(2z'+z)}{(z-z')(z_1-z)}]
\nonumber\\
&I_3(z,z'|\Gamma_1)= q_{3,k}[R_3(z,z')(-\log(-z)
+\log(-z'))+3\frac{(z+z')}{(z-z')}]
\nonumber\\
& R_3(z,z') \equiv \frac{(z^2+4zz'+z'^2)}{(z-z')^2}
\label{i-3}
\end{align}
Note that $I_3$ is essentially obtained from the lower limit of the
integral, i.e. from $- g(0,z,z')$.  The contour $\Gamma_1$ in $I_3$
specifies which branch of the log is to be taken. In particular
\begin{align}
I_3(z,z'|\Gamma_1)-  I_3(z,z'|\tilde\Gamma_1)
= -2\pi i q_{3,k}\ R_3(z,z')
\label{i-3-jump}
\end{align}
In the long time limit \eq{long-time}, we get
\begin{align}
I_3(z,z'|\Gamma_1) =  I_3(z,z'|\tilde\Gamma_1)
= 2 q_{3,k} t (2\pi/\b) +  q_{3,k} \times \hbox{const} + O(e^{-2\pi t/\b})
\label{i-3-long}
\end{align} 
In this equation we have displayed the principal value of the
relevant integrals (the discontinuity \eq{i-3-jump} tells
us the coefficient of the log term or the linear $t$ term).

However, we would like to understand the above results more simply,
by using the $\wc_n(z_1) \varphi_k(z)$ OPE which is of the form:
\begin{align}
\wc_n(z_1)\varphi_k(z)=
 q_{n,k}\frac{\varphi_k(z)}{(z_1-z)^n}
 +\sum_{i=1}^{n-1}\alpha_{n,i} \frac{\varphi_{k,i}(z)}{(z_1-z)^{n-i}}+
\hbox{regular terms}
\label{w-phi-OPE}
\end{align}
where $\varphi_{k,i}(z)$ is of dimension $h_k+i$.\footnote{This is the
  general form; some of the $\a_{n,i}$ coefficients may, of course,
  vanish.} Using this, we get an expansion for the connected 3-point
function of the form:
\begin{align}
&\frac{\lan\wc_n(z_1)\varphi_k(z)\varphi^*_k(z')
\ran_{\mathbb{C}}^{conn}
}{\lan\varphi_k(z)\varphi^*_k(z')\ran_{\mathbb{C}}^{conn}}
=  \frac{q_{n,k}}{(z_1-z)^n} +  \frac{C_{n,1}}
{(z_1-z)^{n-1}(z-z')}
+ O(z-z')^{-2} 
\end{align} 
Performing the integral in \eq{ratio-o-mu-n},
\begin{align}
g_n(z_1,z,z') &= q_{n,k} \left(
\log[z_1-z] - (n-1)\frac{z}{z_1-z} 
+ ...\right)
\nonumber\\&
+ \frac{C_{n,1}}{z-z'}\left( z_1-z +
(n-1) z \log[z_1-z]  + ...\right) + ...
\nonumber
\end{align}
The ellipsis in each round bracket represents terms with higher powers
of $1/(z_1-z)$ (up to a maximum of $(z_1-z)^{-n}$); successive round
brackets themselves are arranged in higher inverse powers of
$z-z'$. Using the $\wc_n(z_1) \varphi^*_k(z')$ OPE in a similar
fashion and using the symmetry property $g_n(z_1,z,z')= (-1)^n
g_n(z_1,z',z)$ we can arrive at a general structure
\begin{align}
g_n(0,z,z') &= q_{n,k} (\log[-z] -\log[-z']) R_n(z,z') + ...
\nonumber
\end{align}
where $R_n(z,z')= (-1)^{n-1}R_n(z',z)$ is of the form
$P_{n-1}(z,z')/(z-z')^{n-1}$ ($P_{n-1}(z,z')$ is a homogeneous
symmetric polynomial of degree zero). See the explicit form of $R_n$
for $n=3$ in \eq{i-3}. The omitted terms are all ratios of homogeneous
polynomials in $(z,z')$ of the same degree in the numerator and in the
denominator. This implies that we have, in the long time limit
\eq{long-time}
\begin{align}
I_n(z,z'|\Gamma_1) =  I_3(z,z'|\tilde\Gamma_1)
= 2 q_{n,k} (2\pi/\b)t +  q_{n,k} \times \hbox{const} + O(e^{-2\pi t/\b})
\label{i-n-long}
\end{align}
which, of course, agrees with \eq{i-3-long}. 

Note that the dominant time-dependence $2 q_{n,k} t (2\pi/\b)$ comes
from the long-time limit of the coefficient $R_n(z,z')$ of the log
terms, which can be read off from the discontinuity
$I_n(z,z'|\Gamma_1) - I_n(z,z'| \tilde\Gamma_1)$ (see \eq{i-3-jump}).
Now, the contour $\int_{\Gamma_1 - \tilde\Gamma_1} dz_1$ can be
deformed to a very small circle $\oint \Gamma_z dz_1$ around the point
$z$; therefore the leading long-time behaviour $R_n^{(0)}(z,z')$ can
be derived by using the leading OPE singularity in \eq{w-phi-OPE} and
computing the residue at $z_1=z$:
\begin{align}
&\hbox{Coefficient of} ~ [\log(-z')- \log(-z)]~{\rm in}~I_n(z,z')
\nonumber\\
&={\hbox{Residue}}_{z_1=z}\left( \frac{\lan\wc_n(z_1)\varphi_k(z)\varphi^*_k(z')
\ran_{\mathbb{C}}^{conn}
}{\lan\varphi_k(z)\varphi^*_k(z')\ran_{\mathbb{C}}^{conn}}
\right)
\equiv q_{n,k} R_n^{(0)}(z,z') =  q_{n,k} 
\label{residue-o-mu}
\end{align}

\section{\label{sec:short}Short interval expansion}

In this section we will explain a formalism suitable for computing
partition functions of the kind that appear in \eq{I-t-zs}. 
For convenience we will first compute these quantities in Euclidean
time $\t = i t$ and later analytically continue back to Lorentzian
time. With this, each of the expressions $Z_{sc}, Z_{ss}, Z_{cc}$ 
is of the form
\begin{align}
\Tr( \rho_{A,1} \rho_{A,2}) = 
\int_{\rm geometry~ 1}\kern-27pt {\mathbf D}\varphi_1 \;\;
\int_{\rm geometry~ 2}\kern-27pt {\mathbf D}\varphi_2\;\;
\delta(F[\varphi_1, \varphi_2])
\exp\left(- S[\varphi_1] -  S[\varphi_2]\right) 
\label{gluing}
\end{align}
where $S[\varphi]$ represents the action for the CFT (with fields
$\varphi$) and the delta-functional in the measure represents a gluing
condition between a geometry `1' and a geometry `2' along a `cut'
which is the location, at a particular time $\t$, of the spatial
interval $A:\s\in (-l/2, l/2)$ \footnote{To be precise, $\delta[F]=$ $
  \delta(\varphi_1(A_<)-\varphi_2(A_>))$
  $\delta(\varphi_1(A_>)-\varphi_2(A_<))$, where $A_<$ ($A_>$)
  represents the limiting value from below (above) the cut.}  For
$Z_{ss}$, both geometries are that of a strip of the Euclidean plane
described by complex coordinates $(w, \bar w)$ $ = \s \pm i \t $
defined by boundaries at $\t = \pm \b/4$ with boundary conditions
determined by the boundary state $| Bd \ran$ introduced in
\eq{psi-0}. For $Z_{cc}$, both geometries are that of a cylinder cut of
the Euclidean plane with identified boundaries at $\t=-\b/4, 3\b/4$.
The geometries for both $Z_{ss}$ and $Z_{cc}$ are familiar from calculations
of Entanglement Renyi entropy (of order 2) and can be calculated from
appropriate correlation functions of twist fields
\cite{Calabrese:2004eu} which exchange two identical geometries. For
$Z_{sc}$, the two glued geometries are different (that of a strip and a
cylinder), hence the method of twist operators do not apply in a
straightforward fashion. (See Figure \ref{tonni-fig}). In this paper,
we will therefore, employ the method of the short interval expansion.

\begin{figure}[H]
\centering
\includegraphics[width=475pt, height=100pt]{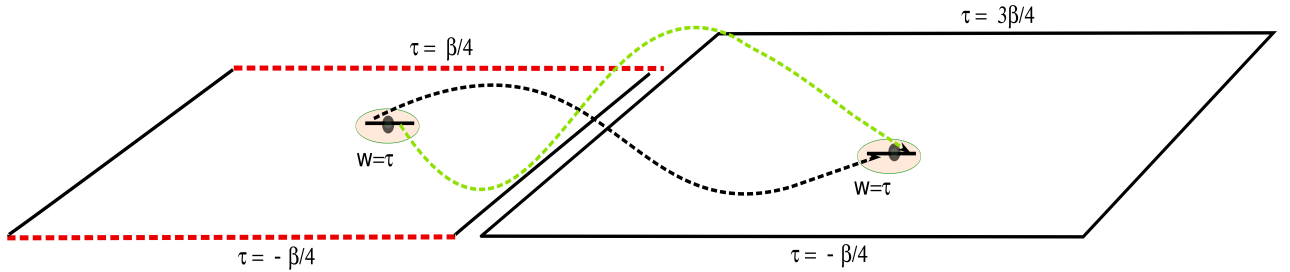}
\caption{\footnotesize Two different geometries, the strip and the
  cylinder, glued along the cut as described in the text. The method
  of the short interval expansion allows us to compute the functional
  integral over this geometry by replacing a small tube enclosing the
  two glued cuts by a complete basis of operators $\phi_{k_1} \otimes
  \phi_{k_2}$ where the operators live in the two Hilbert spaces.}
\label{tonni-fig}
\end{figure}

The idea of the short interval expansion \cite{Calabrese:2010he} is as
follows. To begin, we express the functional integral \eq{gluing} as
an overlap of two wavefunctions in $\mathsf H_1 \otimes \mathsf H_2$,
as follows
\begin{align}
&Z_{12}=\Tr( \rho_{A,1} \rho_{A,2}) = \lan \psi_{out} |  \psi_{in} \ran 
=   \int_{w_1 \in {\cal D}_1}\kern-27pt 
{\mathbf D}\overline\varphi_1(w_1) \kern-5pt
\int_{w_2 \in {\cal D}_2}\kern-27pt {\mathbf D}\overline
\varphi_2(w_2)\;\;
\psi_{in}[\overline{\varphi_1},\overline{\varphi_2}] \;
\psi_{out}^*[\overline{\varphi_1},\overline{\varphi_2}]  
\nonumber\\
& \psi_{in}[\overline{\varphi_1},\overline{\varphi_2}]  
\equiv \int_{w_1 \in {\cal D}_1}\kern-27pt {\mathbf D}\varphi_1(w_1) \kern-5pt
\int_{w_2 \in {\cal D}_2}\kern-27pt {\mathbf D}\varphi_2(w_2)
\delta(\varphi_1|_{\del {\cal D}_1}  - \overline{\varphi_1})
\delta(\varphi_2|_{\del {\cal D}_2}  - \overline{\varphi_2})
\delta(F[\varphi_1, \varphi_2])
\exp\left(- S[\varphi_1] -  S[\varphi_2]\right) 
\nonumber\\
& \psi_{out}[\overline{\varphi_1},\overline{\varphi_2}]  
\equiv \int_{w_1 \notin {\cal D}_1}\kern-27pt {\mathbf D}\varphi_1(w_1)\kern-5pt
\int_{w_2 \notin {\cal D}_2}\kern-27pt {\mathbf D}\varphi_2(w_2)
\delta(\varphi_1|_{\del {\cal D}_1}  - \overline{\varphi_1})
\delta(\varphi_2|_{\del {\cal D}_2}  - \overline{\varphi_2})
\exp\left(- S[\varphi_1] -  S[\varphi_2]\right) 
\label{short-detail}
\end{align}
Here ${\cal D}_1$ (respectively, ${\cal D}_2$) is a small disc drawn
around the cut in geometry 1 (respectively, geometry 2). 

Note that only $| \psi_{in} \ran$ depends on the gluing condition
since the delta functional in the measure does not affect $|
\psi_{out} \ran$.  The basic point of the short interval is that in
the limit when the length $l$ of the cut is small compared with the
characterizing length scale of the geometries (in our case, when $l
\ll \b$), the wavefunction $\psi_{in}[\varphi_1, \varphi_2]$ becomes
jointly localized at the centre $(w_1, \bar w_1)$ of the disc ${\cal
  D}_{1}$ {\it and} at the centre $(w_2, \bar w_2)$ of the disc ${\cal
  D}_{2}$ \footnote{\label{ftnt:midp-pt}We will take the centre of the
  disc in each geometry to coincide with the centre of the cut, which
  has coordinates $w= i\t, \bar w=-i\t$.}, and hence can be expanded in
terms of local operators, as follows
\begin{align}
| \psi_{in} \ran =  \sum_{k_1, k_2} C_{k_1, k_2}~ \phi_{k_1}(w_1, \bar w_1)\,
\phi_{k_2}(w_2, \bar w_2) | 0 \ket_1 \otimes | 0 \ket_2
\label{short-expansion}
\end{align}
Here $k_1, k_2$ label a complete basis of quasiprimary operators of
the CFT Hilbert space. Each term in the sum represents a factorized
wavefunction (between geometries 1 and 2), which, therefore, gives
\footnote{In case geometries 1 and 2 are identical, the superscripts
in $w_i, \wb_i, i=1,2$ indicate which sheet we are considering.}
\begin{align}
& \hat Z_{sc} =\sum_{k_1, k_2} C_{k_1,k_2} \lan \phi_{k_1}(w_1,
\wb_1) \ran_{str} \lan
  \phi_{k_2}(w_2, \wb_2) \ran_{cyl}, \nonumber\\
& \hat Z_{ss} =\sum_{k_1, k_2}
  C_{k_1,k_2} \lan \phi_{k_1}(w_1,\wb_1) \ran_{str} \lan \phi_{k_2}(w_2, \wb_2)
  \ran_{str},\, \nonumber\\
&\hat Z_{cc} =\sum_{k_1, k_2} C_{k_1,k_2} \lan
  \phi_{k_1}(w_1,\wb_1) \ran_{cyl} \lan \phi_{k_2}(w_2, \wb_2) \ran_{cyl}
\label{short-interval}
\end{align}
Here the subscripts $str$ and $cyl$ refer to ``strip'', and
``cylinder'' respectively. The one-point functions are evaluated on
the respective geometries without any cut (see Section
\ref{sec:one-pt} for more details). The glued functional integral
\eq{gluing}, \eq{short-detail} is recovered by summing over $k_1, k_2$
with the coefficients $C_{k_1, k_2}$; , as clear from
\eq{short-interval} these are determined by the gluing condition and
depend on the size of the cut \cite{Calabrese:2010he} (see Section
\ref{app:ck1k2} for more details).

\subsection{\label{app:ck1k2}The coefficients $C_{k_1, k_2}$}

As explained in \cite{Calabrese:2010he} (see also Section
\ref{sec:short}), the coefficients $C_{k_1, k_2}$ are
determined by the equation
\begin{align}
C_{k_1, k_2}= \frac{Z_2}{Z_1^2}({n_{k_1} n_{k_2}})^{-\frac12}\,
\kern-10pt\lim_{z_1 \to \infty_1, z_2 \to \infty_2} \kern-5pt
(z_1 z_2)^{2(h_{k_1}+h_{k_2})}  
(\zb_1 \zb_2)^{2(\hb_{k_1}+\hb_{k_2})} 
\lan \phi_{k_1}(z_1, \zb_1) \phi_{k_2}(z_2, \zb_2) \ran_{{\mathbb C}_2}
\label{ck1k2-def}
\end{align}
where ${\mathbb C}_2$ represents two infinite planes glued along a cut
$A$, $Z_2$ is the functional integral such a glued geometry and $Z_1$
is the functional integral over a single plane. This equation can be
easily proved by inserting quasiprimary a operator at infinity in each
plane in an equation like \eq{gluing} or \eq{short-detail}.  The
two point function in the glued geometry is to be determined by
using the uniformizing map:
\begin{align}
y= \sqrt{(z+l/2)/(z-l/2)}
\label{uniform}
\end{align}
The normalization constants $n_k$ are
determined by the following orthogonality condition of 
the quasiprimary operators
\begin{align}
\lan \phi_{k_1}(z_1, \zb_1) \phi_{k_2}(z_2, \zb_2) \ran_{{\mathbb C}}
=  \frac{n_{k_1}\delta_{k_1, k_2}}{z_{12}^{h_{k_1}+ h_{k_2}}\zb_{12}^{\hb_{k_1}+ \hb_{k_2}}}
\end{align}
where $n_{k_1}$ is a normalization constant. Note that
$C_{k_1, k_2}= C_{k_2, k_1}$. Below we will use the notation
\begin{align}
\hat C_{k_1, k_2}= C_{k_1, k_2}/C_{0,0}
\label{c-hat}
\end{align}

\ni\underbar{Case $(k_1, k_2)=(0,0)$}: We will denote the identity
operator as $\phi_0=1$. It is obvious that
\begin{align}
& C_{0,0}=  Z_2/Z_1^2
\end{align} 
\ni\underbar{Case $(k_1, k_2)=(k,0)$}: The only case where $C_{k,0}
\ne 0$ is when $\phi_k(z,\zb)$ is a descendent of the identity
operator, e.g.  $T(z)$, $\bar T(\zb)$, $\nor{T(z)\bar T(\zb)}$,
$\Lambda(z)$, $\Lambda(\zb)$ etc.\footnote{Here
  $\Lambda(z)=\ \nor{TT}(z) - \frac3{10} \del_z^2T$ is the level 4
  quasiprimary descendent of the identity.} E.g.
\begin{align}
\hat C_{T,0}= C_{T,0}/C_{0,0}= \hat C_{\bar T,0}=  \frac{l^2}{16}; 
\hat C_{T\bar T, 0} =  \frac{l^4}{256}; ... 
\label{c-t}
\end{align}
All other $C_{k,0}$ vanish as they are proportional to 
a one-point function of a primary operator on the Riemann
surface (and hence to that on the complex plane). 

\ni\underbar{Case $(k_1, k_2)=$ (primary, primary)}:
In case $\phi_{k_1}, \phi_{k_2}$ are primary operators, 
\eq{ck1k2-def} gives
\begin{align}
\hat C_{k_1, k_2} = \frac1{n_{k_1}}\delta_{k_1, k_2}
\left(\frac{le^{i\pi/2}}4\right)^{2(h_{k_1}+ \hb_{k_1})} 
\label{ck1k2-val}
\end{align}
\ni\underbar{Case $(k_1, k_2)=$ (descendent, descendent)}: In case
$\phi_{k_1}$ is of the form $L_{-n_1}L_{-n_2}...\bL_{-m_1}
\bL_{-m_2}...\phi_{l_1}$ and $\phi_{k_2}$ is of the form
$L_{-r_1}L_{-r_2}...\bL_{-s_1} \bL_{-s_2}...\phi_{l_2}$, we can show that
\begin{align}
&\hat C_{k_1, k_2} = \delta_{l_1, l_2}\ \delta_{\,\sum n,\sum r}
\ \delta_{\,\sum m, \sum s}\, A(n_1,n_2,..., m_1, m_2,...; r_1, r_2,...,
s_1, s_2,...) \, l^{2(h_{k_1}+ \hb_{k_1})}, 
\nonumber\\ 
& h_{k_1}= h_{l_1} + \sum
n, \;\; h_{k_2} = h_{l_2} + \sum m
\label{descendent}
\end{align}
where $A(...)$ is a numerical coefficient.

\bibliographystyle{jhepmod}
\bibliography{thermal-bib}

\end{document}